\documentclass[useAMS,usenatbib]{mn2e}

\usepackage{graphicx}

\title[The Hyades is chemically inhomogeneous]{The Hyades open cluster is chemically inhomogeneous}

\author[F. Liu et al.]{F. Liu,$^{1}$\thanks{E-mail: fan.liu@anu.edu.au}
D. Yong,$^{1}$
M. Asplund,$^{1}$
I. Ram\'irez,$^{2}$
J. Mel\'endez,$^{3}$\\
$^{1}$Research School of Astronomy and Astrophysics, Australian National University, Canberra, ACT 2611, Australia\\
$^{2}$McDonald Observatory and Department of Astronomy, University of Texas at Austin, 2515 Speedway, Austin, TX 78712-1205, USA\\
$^{3}$Departamento de Astronomia do IAG/USP, Universidade de Sao Paulo, Rua do Matao 1226, Sao Paulo 05508-900, SP, Brasil}

\begin{document}

\date{Accepted 2016 January 27. Received 2016 January 25; in original form 2015 May 20}

\pagerange{\pageref{firstpage}--\pageref{lastpage}} \pubyear{2015}

\maketitle

\label{firstpage}

\begin{abstract}
We present a high-precision differential abundance analysis of 16 solar-type stars in the Hyades open cluster based on high resolution, high signal-to-noise ratio (S/N $\approx$ 350 - 400) spectra obtained from the McDonald 2.7m telescope. We derived stellar parameters and differential chemical abundances for 19 elements (C, O, Na, Mg, Al, Si, S, Ca, Sc, Ti, V, Cr, Mn, Fe, Co, Ni, Cu, Zn, and Ba) with uncertainties as low as $\sim$ 0.01 - 0.02 dex. Our main results include: (1) there is no clear chemical signature of planet formation detected among the sample stars, i.e., no correlations in abundances versus condensation temperature; (2) the observed abundance dispersions are a factor of $\approx$ 1.5 - 2 larger than the average measurement errors for most elements; (3) there are positive correlations, of high statistical significance, between the abundances of at least 90\% of pairs of elements. We demonstrate that none of these findings can be explained by errors due to the stellar parameters. Our results reveal that the Hyades is chemically inhomogeneous at the 0.02 dex level. Possible explanations for the abundance variations include (1) inhomogeneous chemical evolution in the proto-cluster environment, (2) supernova ejection in the proto-cluster cloud, and (3) pollution of metal-poor gas before complete mixing of the proto-cluster cloud. Our results provide significant new constraints on the chemical composition of open clusters and a challenge to the current view of Galactic archeology.
\end{abstract}


\section{Introduction}

Despite decades of studies, we still lack a thorough knowledge of the sequence of events involved in the formation of the Galactic disc \citep{edv93,chi01,ben14,kub15,mg15}. Stellar chemical abundances are expected to keep the fossil record of the conditions of the Galactic disc at the time of its formation. Therefore, careful measurements of stellar chemical abundances using high resolution spectroscopy can reveal the nature of star-forming aggregates, and the detailed chemical and dynamical evolution of the Galactic disc. In the current view of Galactic archeology, star-forming aggregates imprint unique chemical signatures, which can be used to identify and track individual stars back to a common birth site, namely, chemical tagging \citep{fb02}. Such associations would therefore provide key new insights into the early star formation processes. However, several conditions must be met in order for chemical tagging to be successful \citep{bf04,bla10a,bc15,de15}. The pre-requisite is that the open clusters, which are likely the remnants of star-forming aggregates in the Galactic disc, should be chemically homogeneous. The second pre-requisite is that there should be clear cluster-to-cluster abundance differences. Determining the level of chemical homogeneity in open clusters is thus of fundamental importance in the study of the evolution of star-forming clouds and that of the Galactic disc.

Previous studies (e.g., \citealp{fb92,pau03,de06,de07,ti12,fri14}) have argued that open clusters are chemically homogeneous, except for Li \citep{bt86}, Be \citep{smi10}, C and N, as they are affected by stellar evolution, implying that the progenitor cloud was uniformly mixed before its stars formed. However, the observed abundance dispersions are typically $\sim$ 0.05 dex or larger, and can be attributed entirely to the measurement uncertainties. \citet{on11} and \citet{on14} successfully achieved a higher precision level ($\sim$ 0.03 dex) by using strictly differential analysis on the open cluster M67 and found that this rich open cluster has a chemical composition very close to the solar composition, providing significant clues regarding solar birth place, although \citet{pic12} argued against this point based on their dynamical results. Theoretical studies from \citet{bla10b} indicate that a proto-cluster cloud should have sufficient time to homogenize before the first supernova explodes, for clusters with mass of $\sim 10^5 - 10^7$ M$_\odot$. Simulations by \citet{fk14} showed that turbulent mixing could homogenize the elemental abundances of a proto-could and thus create an internal abundance dispersion at least five times more homogeneous than the proto-cluster cloud. \citet{bo16} investigated the abundance spread in open clusters and derive limits on the initial abundance spread of 0.01 - 0.03 dex for different elements. Both observations and theory agree that open clusters less massive than $\sim 10^7$ M$_\odot$ should be chemically homogeneous, except perhaps for the internal abundance trends observed in the light elements of all known globular clusters (e.g., \citealp{kra94}). 

Strictly differential line-by-line analysis for measuring relative chemical abundances in stars with very high precision ($\sim$ 0.01 dex) has been applied to different cases over the past few years \citep{mel09,mel12,dy13,liu14,liu16,ram14a,ram15,tm14,bia15,nis15,saf15,spi16}. This unprecedented precision can help us to reveal minor abundance differences in the photospheres of stars. \citet{mel09} demonstrated that the Sun shows a peculiar chemical pattern when compared to most solar twins, namely, a depletion of refractory elements relative to volatile elements. They tentatively attributed this pattern to the formation of planets, especially rocky planets, in the Solar system. In their scenario, refractory elements in the proto-solar nebula were locked up in the terrestrial planets, while the remaining dust-cleansed gas was accreted on to the Sun. In contrast, the typical solar twin did not form terrestrial planets as efficiently. \citet{cha10} confirmed quantitatively that the depletion of refractory elements in the solar photosphere is possibly due to the depletion of a few Earth masses of rocky material. This scenario, however, has been challenged by \citet{gh10} and \citet{adi14}. They argued that the observed trend between elemental abundances and condensation temperature (T$_{\rm cond}$) could possibly be due to the differences in stellar ages rather than the presence of planets. \citet{nis15} showed apparent abundance-age correlations for 21 solar-twins. This indicates that chemical evolution in the Galactic disc might play an important role in the explanation of the T$_{\rm cond}$ trends. This is also confirmed by \citet{spi16}, who was able to disentangle Galactic chemical evolution (GCE) effects from possible planet effects.

Most stars and their planets form in open clusters \citep{lad03}. However, in contrast to planets detected around field stars, only $\sim$ 11 planets have been found orbiting stars in open clusters \citep{lm07,sa07,qui12,qui14,mei13,bru14,man15} and most of them are giant planets. Nearby open clusters are all younger than $\sim$ 1 Gyr and thus it is intrinsically more difficult to find small planets there. Nevertheless, stars in open clusters share the same age, initial chemical composition and dynamical environment \citep{ran05}, and open clusters offer advantages over field stars for studying planet formation. For example, open clusters provide a more controlled sample and reduce systematic uncertainties arising from age, i.e., the only thing that changes from star to star is the mass. The Hyades open cluster is a close-by bench mark open cluster with intermediate age of $\sim$ 625 - 750 Myr \citep{per98,bh15}. This cluster has been spectroscopically studied before (e.g., \citealp{pau03,de06,cp11,mad13,dut16}). In this paper, we present a strictly line-by-line differential abundance analysis, in order to answer the following fundamental questions: (a) What is the level of abundance dispersions in the Hyades? (b) Is the Hyades still chemically homogeneous if we can achieve a much better precision ($\sim$ 0.01 - 0.02 dex)? (c) Can we distinguish minor abundance differences in the Hyades which can be attributed to the planet formation?

\section{Sample selection and observations}

\begin{figure}
\centering
\includegraphics[width=\columnwidth]{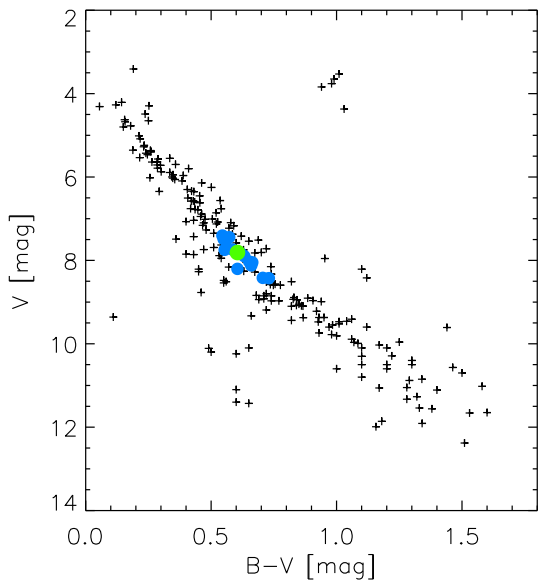}
\caption{Colour-magnitude diagram of the Hyades. The values of B and V magnitude were taken from the Hipparcos Catalog \citep{esa97}. The black plus signs represent 218 Hyades members from \citet{per98} while the blue circles represent our selected programme stars, respectively. The reference star mainly used in this analysis, HD 25825, is the filled green circle.}
\label{fig1}
\end{figure}

\begin{figure}
\centering
\includegraphics[width=\columnwidth]{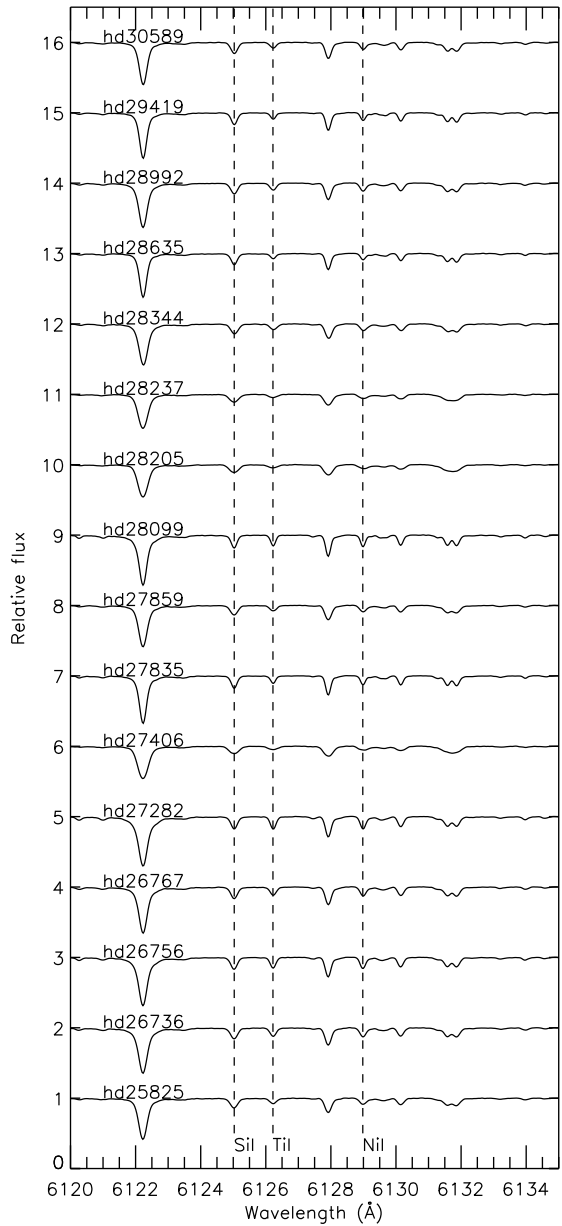}
\caption{A portion of the spectra for all the programme stars. A few atomic lines (Si I, Ti I, Ni I) used in our analysis in this region are marked by the dashed lines.}
\label{fig2}
\end{figure}

We selected 16 solar-type Hyades stars from \citet{pau03} (hereafter P03) with 5650 K $<$ T$_{\rm eff}$ $<$ 6250 K, see Table 1. All of the targets are confirmed Hyades members according to \citet{per98}, except for HD 27835, which was classified as a Hyades member based on its proper motion and radial velocity by \citet{gri88}. According to the SIMBAD data base, 8 sample stars might be variables of BY Draconis type where the variability is caused by star spots. Figure \ref{fig1} shows our selected programme stars in the colour-magnitude diagram. Observations of the targets were performed using the Robert G. Tull Coud\'e Spectrograph \citep{tu95} on the 2.7 m telescope at the McDonald Observatory during two runs in 2012 October and 2012 December. The spectra have a resolving power of R = 60,000 and signal-to-noise ratio (S/N) $\approx$ 350 - 400 per pixel near 6500 \AA. We reduced the spectra with standard procedures which include bias subtraction, flat-fielding, scattered-light subtraction, 1D spectral extraction, wavelength calibration, and continuum normalization, with IRAF\footnote{IRAF is distributed by the National Optical Astronomy Observatory, which is operated by Association of Universities for Research in Astronomy, Inc., under cooperative agreement with National Science Foundation.}. A portion of the reduced spectra for all the programme stars is shown in Figure \ref{fig2}. We note that our S/N ratios are significantly higher than those of P03 who obtained S/N = 100 - 200.

\section{Stellar atmospheric parameters and chemical abundances}

\subsection{Line list}

The line list employed in our analysis was adopted mainly from \citet{sco15a,sco15b} and \citet{gre15} and complemented with additional unblended lines from \citet{ben05} and \citet{nev09}. Equivalent widths (EWs) were measured using the ARES code \citep{sou07}. Weak ($<$ 5 m\AA) and most of strong ($>$ 120 m\AA) lines were excluded from the analysis. The atomic line data, as well as the EW measurements, adopted for the abundance analysis are listed in Table A1. We emphasize that in a strictly line-by-line differential abundance analysis such as ours, the atomic data (e.g., gf values) have essentially no influence on the results since our selected Hyades stars have similar stellar parameters.

\subsection{Establishing parameters for reference stars}

In order to conduct a strictly line-by-line differential analysis, we first need to establish stellar parameters for the reference star(s), and obtained those values in the following manner. We performed a 1D, local thermodynamic equilibrium (LTE) abundance analysis using the 2010 version of MOOG \citep{sne73} with the ODFNEW grid of Kurucz model atmospheres \citep{cas03}. Stellar parameters were obtained by forcing excitation and ionization balance of Fe\,{\sc i} and Fe\,{\sc ii} lines on a line-by-line basis relative to the Sun. The adopted parameters for the Sun were effective temperature (T$_{\rm eff}) = 5777$\,K, surface gravity ($\log g = 4.44$), [Fe/H] = 0.00, and microturbulent velocity ($\xi_{\rm t}$) = 1.00 km\,s$^{-1}$.

We established the stellar parameters of our sample stars by using an automatic grid searching method described by \citet{liu14}. The best combination of T$_{\rm eff}$, $\log g$, [Fe/H] and $\xi_{\rm t}$ were obtained by minimizing the slopes in [Fe\,{\sc i}/H] versus lower excitation potential (LEP) and reduced EW as well as the difference between [Fe\,{\sc i}/H] and [Fe\,{\sc ii}/H], from a successively refined grid of stellar atmospheric models. The final solution was adopted when the grid step-size decreased to $\Delta T_{\rm eff} = 1$ K, $\Delta \log g = 0.01$ and $\Delta \xi_{\rm t}$ = 0.01 km\,s$^{-1}$. We also required the derived average [Fe/H] to be consistent with the adopted model atmospheric value. Lines whose abundances departed from the average by $> 2.5\sigma$ were clipped and the parameters were re-computed after the sigma clipping. Note that if a given line is excluded in one star, the same line is also excluded in all stars. The procedure was applied to all the sample stars since we wanted to be able to select any star as the reference.

\begin{figure}
\centering
\includegraphics[width=\columnwidth]{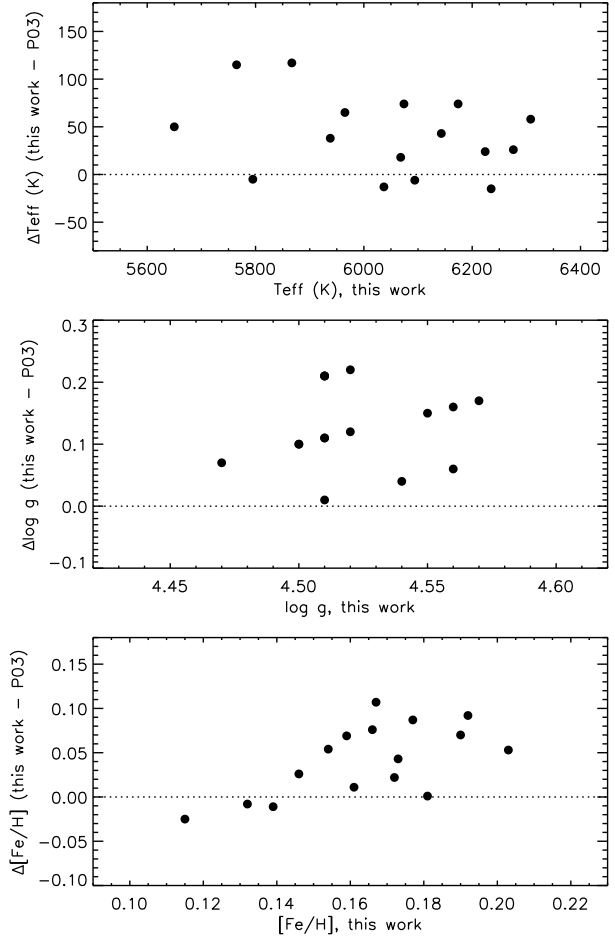}
\caption{Comparison of stellar parameters (top panel: T$_{\rm eff}$; middle panel: $\log g$; bottom panel: [Fe/H]) determined by this work and the study by P03.}
\label{fig3}
\end{figure}

Table 1 lists the stellar atmospheric parameters of our sample stars with the Sun as the reference star. The uncertainties in the stellar parameters were derived with the method described by \citet{eps10} and \citet{ben14}, which accounts for the co-variances between changes in the stellar parameters. We compared our derived stellar parameters with the previous study by P03 in Figure \ref{fig3}. We found that our T$_{\rm eff}$ values follow the one-to-one relation, when compared to P03 results, while this is not the case for $\log g$ and [Fe/H]. The mean differences in T$_{\rm eff}$, $\log g$ and [Fe/H] between our results and P03 results are $41.4 \pm 41.5$ K, $0.13 \pm 0.07$, and $0.04 \pm 0.04$, respectively. We note that we obtained smaller errors in stellar parameters when compared to the results from P03. The errors in our stellar parameters are $\sigma$T$_{\rm eff}$ $\approx$ 28 K, $\sigma$$\log g$ $\approx$ 0.04, $\sigma$[Fe/H] $\approx$ 0.02 and $\sigma$$\xi_{\rm t}$ $\approx$ 0.04 km s$^{-1}$, while the typical errors in stellar parameters from P03 are $\sigma$T$_{\rm eff}$ $\sim$ 50 K, $\sigma$$\log g$ $\sim$ 0.2, $\sigma$[Fe/H] $\sim$ 0.05 and $\sigma$$\xi_{\rm t}$ $\sim$ 0.2 km s$^{-1}$.

\begin{figure}
\centering
\includegraphics[width=\columnwidth]{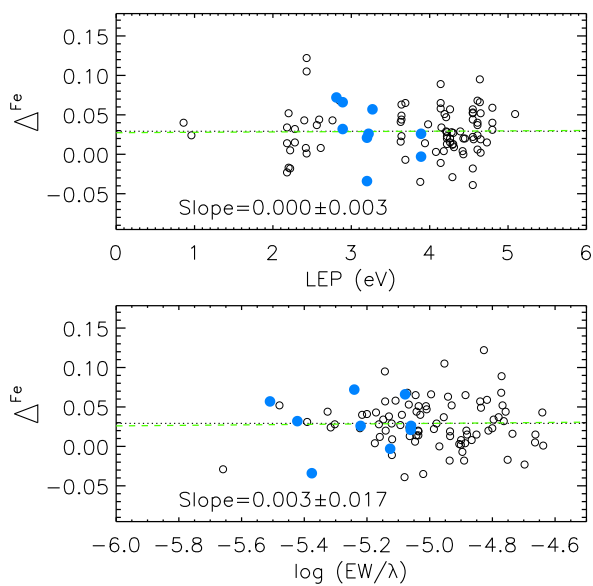}
\caption{Upper panel: differential iron abundances ($\Delta^{\rm Fe}$) of a Hyades star HD 26736 derived on a line-by-line basis with respect to the reference star HD 25825 as a function of LEP; open circles and blue filled circles represent Fe \,{\sc i} and Fe \,{\sc ii} lines, respectively. The black dotted line shows the location of mean $\Delta^{\rm Fe}$, the green dashed line represents the best fit to the data. Lower panel: same as in the top panel but as a function of reduced EW.}
\label{fig4}
\end{figure}

Following \citet{liu14,liu16}, we then derived the differential stellar parameters using a strictly line-by-line differential analysis as described above, but compared our programme stars to a selected reference star from our Hyades sample. Choosing a typical Hyades star as the reference can help us to avoid potential systematic errors arising from comparing the higher metallicity Hyades stars with the Sun. HD 25825, with T$_{\rm eff}$ close to the median value, was selected as the reference star. The adopted stellar parameters for this reference star, T$_{\rm eff} = 6094$\,K, $\log g = 4.56$, [Fe/H] = 0.14, and $\xi_{\rm t}$ = 1.34 km\,s$^{-1}$, were taken from the analysis relative to the Sun (values can be found in Table 1). We emphasize that the absolute values are not crucial for our differential abundance analysis. Figure \ref{fig4} shows an example of determining the differential stellar parameters of a programme star (HD 26736) relative to the reference star HD 25825. The line-by-line differential Fe abundance ($\Delta^{\rm Fe}$) is defined as below. We adopt the notation from \citet{mel12} and \citet{dy13}, the abundance difference (programme star $-$ reference star) for a line is
\begin{equation}
\delta A_i = A^{\rm program \ star}_i - A^{\rm reference \ star}_i
\end{equation}
Therefore, $\Delta^{\rm Fe}$ is
\begin{equation}
\Delta^{\rm Fe} = \ <\delta A^{\rm Fe}_i> \ = \frac{1}{N}\sum_{i=1}^N \delta A^{\rm Fe}_i
\end{equation}
The adopted stellar parameters satisfy the excitation and ionization balance in a differential sense. The best fit $\pm$ 1$\sigma$ for $\Delta^{\rm Fe}$ versus LEP roughly corresponds to an error in T$_{\rm eff}$ of 30 K, similarly for the reduced EW ($\log$ (EW/$\lambda$), which corresponds to an error of $\sim$ 0.03 - 0.04 km\,s$^{-1}$ in $\xi_{\rm t}$. The abundance difference in Fe \,{\sc i} and Fe \,{\sc ii} = 0.000 $\pm$ 0.012, which constrains $\log g$ to a precision of 0.02 - 0.04. The final adopted differential stellar parameters and corresponding errors of our Hyades stars are listed in Table 2. Excellent precision in stellar parameters was achieved due to the strictly line-by-line differential analysis technique, which greatly reduces the systematic errors from atomic line data and shortcomings in the 1D LTE modelling of the stellar atmospheres and spectral line formation (e.g., \citealp{asp05,asp09}).

\begin{table}
\caption{Stellar atmospheric parameters for the programme stars with the Sun as the reference star.}
\begin{tabular*}{\columnwidth}{@{}lcccc@{}}
\hline
Object & T$_{\rm eff}$ (K) & $\log g$ & [Fe/H] & $\xi_{\rm t}$ (km/s) \\
HD 25825 & 6094$\pm$32 & 4.56$\pm$0.05 & 0.139$\pm$0.019 & 1.34$\pm$0.04 \\
HD 26736 & 5867$\pm$31 & 4.50$\pm$0.04 & 0.166$\pm$0.017 & 1.31$\pm$0.04 \\
HD 26756 & 5765$\pm$30 & 4.54$\pm$0.04 & 0.167$\pm$0.015 & 1.17$\pm$0.05 \\
HD 26767 & 5938$\pm$25 & 4.55$\pm$0.04 & 0.190$\pm$0.014 & 1.30$\pm$0.04 \\
HD 27282 & 5650$\pm$28 & 4.51$\pm$0.04 & 0.172$\pm$0.015 & 1.19$\pm$0.05 \\
HD 27406 & 6224$\pm$37 & 4.51$\pm$0.05 & 0.161$\pm$0.024 & 1.42$\pm$0.05 \\
HD 27835 & 6068$\pm$24 & 4.52$\pm$0.03 & 0.177$\pm$0.013 & 1.27$\pm$0.03 \\
HD 27859 & 6037$\pm$27 & 4.51$\pm$0.03 & 0.115$\pm$0.016 & 1.33$\pm$0.04 \\
HD 28099 & 5795$\pm$24 & 4.47$\pm$0.04 & 0.154$\pm$0.016 & 1.22$\pm$0.03 \\
HD 28205 & 6308$\pm$36 & 4.51$\pm$0.05 & 0.192$\pm$0.023 & 1.38$\pm$0.04 \\
HD 28237 & 6235$\pm$37 & 4.51$\pm$0.05 & 0.132$\pm$0.023 & 1.39$\pm$0.05 \\
HD 28344 & 6074$\pm$29 & 4.57$\pm$0.04 & 0.181$\pm$0.019 & 1.29$\pm$0.04 \\
HD 28635 & 6276$\pm$25 & 4.52$\pm$0.03 & 0.159$\pm$0.015 & 1.33$\pm$0.03 \\
HD 28992 & 5965$\pm$22 & 4.51$\pm$0.03 & 0.146$\pm$0.012 & 1.31$\pm$0.03 \\
HD 29419 & 6174$\pm$23 & 4.56$\pm$0.04 & 0.173$\pm$0.013 & 1.32$\pm$0.03 \\
HD 30589 & 6143$\pm$22 & 4.50$\pm$0.03 & 0.203$\pm$0.015 & 1.27$\pm$0.03 \\
\hline
\end{tabular*}
\end{table}

\begin{table}
\caption{Stellar atmospheric parameters for the programme stars relative to a reference star (HD 25825).}
\begin{tabular*}{\columnwidth}{@{}lcccc@{}}
\hline
Object & T$_{\rm eff}$ (K) & $\log g$ & [Fe/H] & $\xi_{\rm t}$ (km/s) \\
\hline
HD 25825$^{a}$ & 6094 & 4.56 & 0.14 & 1.34 \\
\hline
HD 26736 & 5896$\pm$26 & 4.52$\pm$0.04 & 0.168$\pm$0.015 & 1.37$\pm$0.03 \\
HD 26756 & 5760$\pm$24 & 4.54$\pm$0.03 & 0.163$\pm$0.015 & 1.17$\pm$0.04 \\
HD 26767 & 5944$\pm$16 & 4.56$\pm$0.02 & 0.189$\pm$0.008 & 1.31$\pm$0.02 \\
HD 27282 & 5654$\pm$26 & 4.51$\pm$0.04 & 0.172$\pm$0.017 & 1.20$\pm$0.05 \\
HD 27406 & 6225$\pm$32 & 4.51$\pm$0.04 & 0.159$\pm$0.017 & 1.43$\pm$0.04 \\
HD 27835 & 6070$\pm$22 & 4.53$\pm$0.03 & 0.174$\pm$0.013 & 1.28$\pm$0.03 \\
HD 27859 & 6034$\pm$21 & 4.51$\pm$0.03 & 0.111$\pm$0.012 & 1.34$\pm$0.03 \\
HD 28099 & 5819$\pm$28 & 4.49$\pm$0.04 & 0.161$\pm$0.016 & 1.26$\pm$0.04 \\
HD 28205 & 6306$\pm$29 & 4.51$\pm$0.04 & 0.189$\pm$0.015 & 1.39$\pm$0.04 \\
HD 28237 & 6238$\pm$31 & 4.51$\pm$0.04 & 0.130$\pm$0.017 & 1.40$\pm$0.04 \\
HD 28344 & 6074$\pm$16 & 4.57$\pm$0.02 & 0.180$\pm$0.010 & 1.30$\pm$0.02 \\
HD 28635 & 6278$\pm$26 & 4.53$\pm$0.03 & 0.156$\pm$0.012 & 1.34$\pm$0.03 \\
HD 28992 & 5968$\pm$21 & 4.52$\pm$0.03 & 0.143$\pm$0.011 & 1.32$\pm$0.03 \\
HD 29419 & 6180$\pm$25 & 4.57$\pm$0.04 & 0.171$\pm$0.013 & 1.34$\pm$0.03 \\
HD 30589 & 6142$\pm$24 & 4.50$\pm$0.03 & 0.201$\pm$0.011 & 1.27$\pm$0.03 \\
\hline
\end{tabular*}
$^a$ Adopted stellar parameters for the reference star, taken from Table 1.
\end{table}

\subsection{Differential chemical abundances}

Having established the stellar parameters relative to the selected reference star (HD 25825), we derived differential chemical abundances for 19 elements: C, O, Na, Mg, Al, Si, S, Ca, Sc, Ti, V, Cr, Mn, Fe, Co, Ni, Cu, Zn, and Ba with a strictly line-by-line basis. Hyperfine structure splitting was considered for Sc, V, Cr, Mn, Cu, and Ba using the data from \citep{kur95}. NLTE effects in the differential abundance analysis between similar stars such as our sample should be negligible \citep{mel12,mon13}. However, we still applied differential NLTE corrections for the O I triplet using the NLTE calculations by \citet{ama16}. We note that photospheric inhomogeneities caused by star spots might induce differential NLTE or 3D effects on the differential abundances. For example, \citet{mm04} showed that the discrepancy between oxygen abundances derived from the forbidden line at 6300 \AA\, and the O I triplet, increase with increasing chromospheric activity. Such finding could imply that NLTE corrections to the oxygen abundances might not be completely adequate for active stars.

The total error in the differential abundance is obtained by adding in quadrature the standard error of the mean and the errors introduced by the uncertainties in stellar atmospheric parameters following the method of \citet{eps10} which includes co-variance terms. For elements that only one spectral line was measured (C and Zn), we estimate the uncertainties by taking into consideration errors due to S/N, continuum setting and the stellar parameters. The quadratic sum of the three uncertainties sources give the errors for these two elements. Tables 3 - 5 list the differential elemental abundances for our programme stars relative to the reference star HD 25825\footnote{As described before, we define the line-by-line differential abundance for any species, X in this example, as $\Delta^{\rm X}$.}. The precision of the abundance ratios is $\sim$ 0.01 - 0.03 dex for most elements. We note that the strictly line-by-line differential analysis greatly reduces the abundance errors from atomic data and shortcomings in the 1D LTE modelling of the stellar atmospheres and spectral line formation.

\begin{table*}
\caption{Differential abundances $\Delta^{\rm X}$ (C, O, Na, Mg, Al, Si, S, Ca) for our Hyades stars relative to the reference star HD 25825.}
\begin{tabular*}{\textwidth}{@{}lcccccccc@{}}
\hline
Object & $\Delta^{\rm C}$ & $\Delta^{\rm O}$ & $\Delta^{\rm Na}$ & $\Delta^{\rm Mg}$ & $\Delta^{\rm Al}$ & $\Delta^{\rm Si}$ & $\Delta^{\rm S}$ & $\Delta^{\rm Ca}$ \\
\hline
HD 26736 & -0.056$\pm$0.027 & -0.011$\pm$0.030 &  0.038$\pm$0.007 &  0.039$\pm$0.030 &  0.061$\pm$0.033 &  0.038$\pm$0.009 & 0.087$\pm$0.029 &  0.042$\pm$0.016 \\
HD 26756 & -0.041$\pm$0.027 &  0.017$\pm$0.016 &  0.035$\pm$0.019 &  0.021$\pm$0.042 &  0.063$\pm$0.023 &  0.029$\pm$0.006 & 0.026$\pm$0.064 &  0.034$\pm$0.015 \\
HD 26767 & -0.003$\pm$0.026 &  0.050$\pm$0.011 &  0.041$\pm$0.004 &  0.065$\pm$0.013 &  0.073$\pm$0.017 &  0.050$\pm$0.005 & 0.063$\pm$0.014 &  0.054$\pm$0.011 \\
HD 27282 & -0.055$\pm$0.029 &  0.050$\pm$0.030 &  0.053$\pm$0.037 &  0.061$\pm$0.025 &  0.088$\pm$0.016 &  0.043$\pm$0.008 & 0.093$\pm$0.039 &  0.049$\pm$0.011 \\
HD 27406 & -0.040$\pm$0.026 &  0.029$\pm$0.020 &  0.028$\pm$0.019 & -0.052$\pm$0.025 &  0.020$\pm$0.015 &  0.037$\pm$0.009 & 0.039$\pm$0.014 &  0.014$\pm$0.015 \\
HD 27835 & -0.029$\pm$0.026 &  0.025$\pm$0.016 &  0.019$\pm$0.028 &  0.045$\pm$0.029 &  0.055$\pm$0.045 &  0.031$\pm$0.008 & 0.079$\pm$0.037 &  0.037$\pm$0.017 \\
HD 27859 & -0.074$\pm$0.026 &  0.010$\pm$0.014 & -0.010$\pm$0.019 & -0.007$\pm$0.009 & -0.026$\pm$0.023 & -0.023$\pm$0.006 & 0.056$\pm$0.051 & -0.018$\pm$0.013 \\
HD 28099 &  0.003$\pm$0.027 &  0.019$\pm$0.023 &  0.038$\pm$0.019 &  0.050$\pm$0.040 &  0.081$\pm$0.038 &  0.026$\pm$0.011 & 0.071$\pm$0.023 &  0.033$\pm$0.020 \\
HD 28205 & -0.032$\pm$0.026 &  0.054$\pm$0.012 &  0.018$\pm$0.021 &  0.022$\pm$0.042 &  0.027$\pm$0.027 &  0.069$\pm$0.008 & 0.087$\pm$0.038 &  0.031$\pm$0.011 \\
HD 28237 & -0.077$\pm$0.026 & -0.026$\pm$0.018 & -0.024$\pm$0.012 & -0.039$\pm$0.017 & -0.087$\pm$0.061 &  0.015$\pm$0.010 & 0.021$\pm$0.049 & -0.018$\pm$0.015 \\
HD 28344 & -0.027$\pm$0.026 &  0.028$\pm$0.022 &  0.021$\pm$0.005 &  0.033$\pm$0.007 &  0.055$\pm$0.013 &  0.041$\pm$0.006 & 0.058$\pm$0.033 &  0.041$\pm$0.010 \\
HD 28635 & -0.039$\pm$0.026 &  0.004$\pm$0.011 &  0.018$\pm$0.033 &  0.029$\pm$0.029 &  0.005$\pm$0.045 &  0.026$\pm$0.009 & 0.049$\pm$0.050 &  0.020$\pm$0.016 \\
HD 28992 & -0.037$\pm$0.026 &  0.028$\pm$0.019 &  0.025$\pm$0.010 &  0.009$\pm$0.011 &  0.033$\pm$0.027 &  0.020$\pm$0.007 & 0.005$\pm$0.006 &  0.005$\pm$0.010 \\
HD 29419 & -0.032$\pm$0.026 &  0.006$\pm$0.020 &  0.035$\pm$0.030 &  0.044$\pm$0.028 &  0.030$\pm$0.035 &  0.038$\pm$0.008 & 0.077$\pm$0.038 &  0.027$\pm$0.014 \\
HD 30589 & -0.060$\pm$0.026 &  0.019$\pm$0.015 &  0.055$\pm$0.020 &  0.062$\pm$0.019 &  0.070$\pm$0.032 &  0.075$\pm$0.007 & 0.082$\pm$0.053 &  0.060$\pm$0.018 \\
\hline
\end{tabular*}
\end{table*}

\begin{table*}
\caption{Differential abundances $\Delta^{\rm X}$ (Sc, TiI, TiII, V, CrI, CrII, Mn, Fe) for our Hyades stars relative to the reference star HD 25825.}
\begin{tabular*}{\textwidth}{@{}lcccccccc@{}}
\hline
Object & $\Delta^{\rm Sc}$ & $\Delta^{\rm TiI}$ & $\Delta^{\rm TiII}$ & $\Delta^{\rm V}$ & $\Delta^{\rm CrI}$ & $\Delta^{\rm CrII}$ & $\Delta^{\rm Mn}$ & $\Delta^{\rm Fe}$ \\
\hline
HD 26736 &  0.018$\pm$0.023 &  0.048$\pm$0.017 & -0.025$\pm$0.020 &  0.044$\pm$0.0236 &  0.034$\pm$0.017 &  0.023$\pm$0.033 &  0.051$\pm$0.018 &  0.029$\pm$0.010 \\
HD 26756 &  0.034$\pm$0.021 &  0.048$\pm$0.017 & -0.008$\pm$0.032 &  0.026$\pm$0.0320 &  0.031$\pm$0.015 &  0.032$\pm$0.032 &  0.045$\pm$0.013 &  0.024$\pm$0.009 \\
HD 26767 &  0.036$\pm$0.015 &  0.065$\pm$0.011 &  0.033$\pm$0.018 &  0.047$\pm$0.0239 &  0.055$\pm$0.010 &  0.060$\pm$0.017 &  0.060$\pm$0.009 &  0.050$\pm$0.006 \\
HD 27282 &  0.049$\pm$0.026 &  0.057$\pm$0.018 & -0.004$\pm$0.030 &  0.044$\pm$0.0366 &  0.043$\pm$0.017 &  0.037$\pm$0.036 &  0.057$\pm$0.019 &  0.032$\pm$0.009 \\
HD 27406 &  0.018$\pm$0.036 &  0.020$\pm$0.023 &  0.001$\pm$0.030 &  0.013$\pm$0.0302 &  0.004$\pm$0.018 &  0.014$\pm$0.024 &  0.046$\pm$0.020 &  0.020$\pm$0.013 \\
HD 27835 &  0.011$\pm$0.031 &  0.054$\pm$0.014 &  0.038$\pm$0.017 &  0.011$\pm$0.0232 &  0.039$\pm$0.012 &  0.043$\pm$0.015 &  0.028$\pm$0.011 &  0.036$\pm$0.009 \\
HD 27859 & -0.054$\pm$0.020 & -0.024$\pm$0.014 & -0.043$\pm$0.012 & -0.029$\pm$0.0165 & -0.030$\pm$0.012 & -0.020$\pm$0.010 & -0.031$\pm$0.010 & -0.028$\pm$0.008 \\
HD 28099 & -0.012$\pm$0.030 &  0.042$\pm$0.021 &  0.015$\pm$0.018 &  0.021$\pm$0.0277 &  0.029$\pm$0.018 &  0.035$\pm$0.035 &  0.043$\pm$0.019 &  0.022$\pm$0.012 \\
HD 28205 &  0.047$\pm$0.039 &  0.049$\pm$0.020 &  0.031$\pm$0.027 &  0.034$\pm$0.0382 &  0.028$\pm$0.018 &  0.029$\pm$0.022 &  0.050$\pm$0.016 &  0.050$\pm$0.011 \\
HD 28237 & -0.035$\pm$0.028 & -0.016$\pm$0.020 & -0.050$\pm$0.032 & -0.007$\pm$0.0267 & -0.023$\pm$0.018 & -0.045$\pm$0.021 &  0.003$\pm$0.023 & -0.008$\pm$0.013 \\
HD 28344 &  0.021$\pm$0.024 &  0.062$\pm$0.012 &  0.014$\pm$0.018 &  0.033$\pm$0.0186 &  0.027$\pm$0.010 &  0.042$\pm$0.011 &  0.038$\pm$0.010 &  0.041$\pm$0.007 \\
HD 28635 & -0.017$\pm$0.032 &  0.032$\pm$0.017 &  0.032$\pm$0.016 & -0.033$\pm$0.0256 &  0.008$\pm$0.014 &  0.014$\pm$0.017 & -0.005$\pm$0.020 &  0.018$\pm$0.010 \\
HD 28992 & -0.022$\pm$0.024 &  0.008$\pm$0.013 & -0.015$\pm$0.016 & -0.014$\pm$0.0183 &  0.006$\pm$0.012 &  0.018$\pm$0.015 &  0.018$\pm$0.014 &  0.004$\pm$0.008 \\
HD 29419 &  0.004$\pm$0.036 &  0.054$\pm$0.017 &  0.034$\pm$0.017 &  0.006$\pm$0.0236 &  0.033$\pm$0.015 &  0.010$\pm$0.017 &  0.026$\pm$0.017 &  0.032$\pm$0.010 \\
HD 30589 &  0.047$\pm$0.022 &  0.075$\pm$0.015 &  0.060$\pm$0.014 &  0.021$\pm$0.0196 &  0.061$\pm$0.012 &  0.049$\pm$0.013 &  0.060$\pm$0.010 &  0.062$\pm$0.009 \\
\hline
\end{tabular*}
\end{table*}

\begin{table*}
\caption{Differential abundances $\Delta^{\rm X}$ (Co, Ni, Cu, Zn, Ba) for our Hyades stars relative to the reference star HD 25825.}
\begin{tabular}{@{}lccccc@{}}
\hline
Object & $\Delta^{\rm Co}$ & $\Delta^{\rm Ni}$ & $\Delta^{\rm Cu}$ & $\Delta^{\rm Zn}$ & $\Delta^{\rm Ba}$ \\
\hline
HD 26736 &  0.044$\pm$0.026 &  0.046$\pm$0.014 &  0.058$\pm$0.050 &  0.038$\pm$0.028 & -0.016$\pm$0.015 \\
HD 26756 &  0.017$\pm$0.020 &  0.035$\pm$0.013 &  0.102$\pm$0.072 &  0.087$\pm$0.027 & -0.007$\pm$0.020 \\
HD 26767 &  0.046$\pm$0.011 &  0.049$\pm$0.009 &  0.077$\pm$0.034 &  0.077$\pm$0.026 &  0.058$\pm$0.012 \\
HD 27282 &  0.040$\pm$0.031 &  0.044$\pm$0.013 &  0.037$\pm$0.029 &  0.118$\pm$0.030 &  0.029$\pm$0.018 \\
HD 27406 &  0.009$\pm$0.038 & -0.004$\pm$0.019 &  0.021$\pm$0.042 &  0.001$\pm$0.028 &  0.002$\pm$0.027 \\
HD 27835 &  0.029$\pm$0.024 &  0.042$\pm$0.013 &  0.030$\pm$0.034 &  0.062$\pm$0.027 &  0.022$\pm$0.012 \\
HD 27859 & -0.040$\pm$0.014 & -0.028$\pm$0.012 & -0.025$\pm$0.010 &  0.007$\pm$0.026 & -0.052$\pm$0.010 \\
HD 28099 &  0.013$\pm$0.045 &  0.033$\pm$0.016 &  0.068$\pm$0.068 &  0.087$\pm$0.028 & -0.014$\pm$0.018 \\
HD 28205 &  0.021$\pm$0.029 &  0.037$\pm$0.016 &  0.040$\pm$0.031 &  0.018$\pm$0.028 &  0.009$\pm$0.018 \\
HD 28237 & -0.039$\pm$0.030 & -0.019$\pm$0.018 & -0.025$\pm$0.019 & -0.051$\pm$0.028 & -0.056$\pm$0.020 \\
HD 28344 &  0.038$\pm$0.013 &  0.038$\pm$0.010 &  0.050$\pm$0.023 &  0.063$\pm$0.026 &  0.045$\pm$0.009 \\
HD 28635 &  0.020$\pm$0.031 &  0.021$\pm$0.014 &  0.009$\pm$0.027 & -0.011$\pm$0.027 & -0.011$\pm$0.026 \\
HD 28992 & -0.020$\pm$0.015 &  0.000$\pm$0.012 &  0.000$\pm$0.031 &  0.029$\pm$0.027 & -0.008$\pm$0.013 \\
HD 29419 &  0.049$\pm$0.031 &  0.045$\pm$0.014 &  0.041$\pm$0.021 &  0.005$\pm$0.027 & -0.021$\pm$0.015 \\
HD 30589 &  0.050$\pm$0.030 &  0.073$\pm$0.012 &  0.069$\pm$0.027 &  0.075$\pm$0.027 &  0.010$\pm$0.015 \\
\hline
\end{tabular}
\end{table*}

We repeated the procedure by using each programme star as a reference star and determined the corresponding differential stellar parameters and chemical abundances. We note that changing the reference star does not alter our results and conclusions in general.

\section{Results and discussions}

\subsection{Chemical signatures of planet formation}

\citet{mel09} performed the first high precision differential abundance analysis of the Sun and solar twins and found that the Sun was chemically unusual when compared to the solar twins. They found a clear correlation between abundance differences (Sun $-$ solar twins) as a function of condensation temperature (T$_{\rm cond}$) and suggested that this was related to terrestrial planet formation in the early solar environment. Therefore, we investigate whether the chemical signatures of planet formation can be found in our Hyades stars since identifying planets in open clusters is important to test whether the frequency is the same as in field stars and whether there is any dependence of planet frequency on stellar mass (e.g., \citealp{coc02}). While our programme stars do not host hot Jupiters \citep{pau04}, we do not yet know whether they host smaller planets. 

\begin{figure}
\centering
\includegraphics[width=\columnwidth]{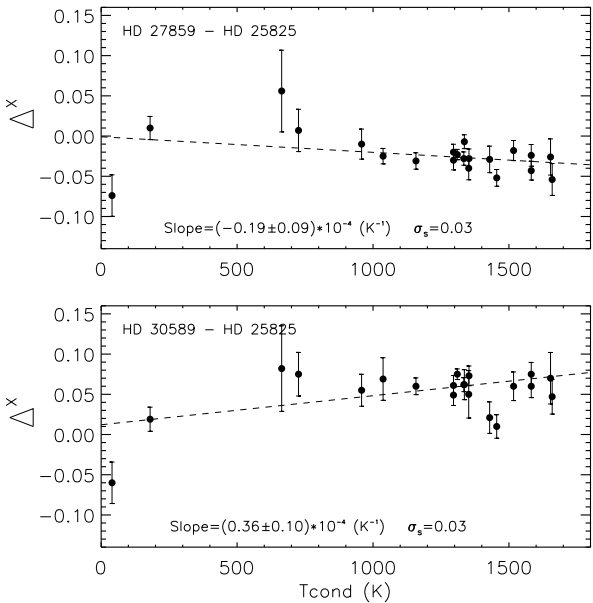}
\caption{Differences in chemical abundances ($\Delta^{\rm X}$) versus condensation temperature (T$_{\rm cond}$) for two programme stars relative to the reference star HD 25825 with the most negative slope (HD 27859, upper panel) and most positive slope (HD 30589, lower panel). The dashed lines represent the linear least-squares fits to the data with the respective slopes given in each panel. $\sigma_s$ is the dispersion about the linear fit.}
\label{fig5}
\end{figure}

\begin{figure}
\centering
\includegraphics[width=\columnwidth]{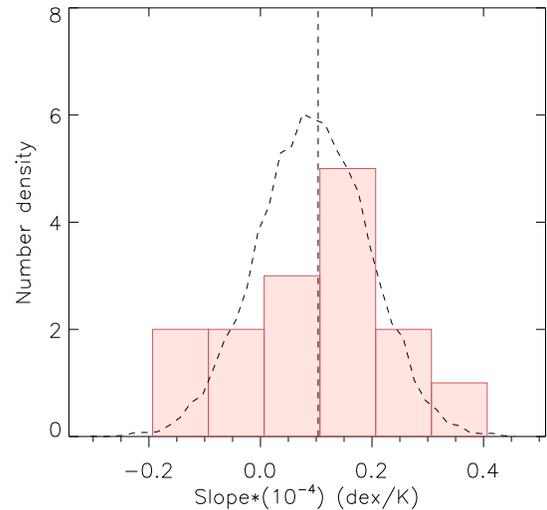}
\caption{Histogram of the slopes when applying a single linear fit to $\Delta^{\rm X}$ - T$_{\rm cond}$ for our Hyades stars relative to the reference star HD 25825. The dashed vertical line represents the location of the mean value of $\Delta^{\rm X}$ versus T$_{\rm cond}$ slopes. The dashed curve represents the distribution of slopes of data with pure observational noise (see text for details).}
\label{fig6}
\end{figure}

With our selected reference star (HD 25825), we obtained the differential chemical abundance ($\Delta^{\rm X}$) versus T$_{\rm cond}$ relations for each programme star; T$_{\rm cond}$ were taken from \citet{lod03}. In Figure \ref{fig5}, we show two sample stars (HD 27859 and HD 30589) with largest, and smallest depletion in refractory elements compared to the reference star (i.e., most negative, and most positive slope, respectively). For HD 27859, the amplitude of depletion is only $\approx$ 0.03 dex and the significance level of the slope is 2$\sigma$. For HD 30589, the amplitude of enrichment is $\approx$ 0.07 dex and the significance level of the slope is 3.6$\sigma$. If the hypothesis suggested by \citet{mel09} is true, HD 27859 might have higher chance to host a terrestrial planet(s) due to the depletion pattern in refractory elements. However, the low value of $\Delta^{\rm C}$, and the low statistical significance make it hard to draw such a conclusion. We show the histogram of the slopes for the single linear fit to the T$_{\rm cond}$ trends for our Hyades stars in Figure \ref{fig6}. The slopes exhibit a broad distribution with a mean of $\sim$ 0.11 $\times$ 10$^{-4}$ K$^{-1}$. We did not find any programme stars with a clear chemical pattern with high significance. Following \citet{ram14b}, we generated 10,000 $\Delta^{\rm X}$ versus T$_{\rm cond}$ relations, with the $\Delta^{\rm X}$ values drawn from a Gaussian distribution of 0.02 dex of standard deviation (this corresponds to the typical abundance errors in our analysis) centred at zero. We calculated the $\Delta^{\rm X}$ versus T$_{\rm cond}$ slopes for each of these relations and determined their distribution, namely "trial distribution", normalized to have an equal area to the number of programme stars in our real sample. We shift the mean of the "trial distribution" to the mean of T$_{\rm cond}$ slopes of our data and over-plot this "trial distribution" in Figure \ref{fig6}. We note that the "trial distribution" has a width very similar to the real distribution of our data. We applied the Kolmogorov-Smirnov (K-S) test to compare the shifted "trial distribution" and the real distribution of our data. We obtained the D-value of $\approx$ 0.2 and the p-value of $\approx$ 0.5. This further demonstrates that in fact there are no T$_{\rm cond}$ correlations in our data.

We have repeated the analysis using each programme star in turn as the reference star. We find no clear T$_{\rm cond}$ trends which might indicate the chemical signature of planet formation in our sample stars. 

We note that \citet{qui14} detected one hot Jupiter around a Hyades open cluster star and they suggested a hot Jupiter frequency of  $1.97^{+0.92}_{-1.07}$\% in the Hyades open cluster, which is consistent with the hot Jupiter frequency in the field stars (1.2$\pm$ 0.38\%, \citealp{wri12}), while no hot Jupiters were discovered around our selected Hyades stars \citep{pau04}. \citet{mei13} detected two planets smaller than Neptunes around two Sun-like stars in the old open cluster NGC 6811 and argued that the small planet frequency in the open cluster stars is the same as the frequency in the filed stars. \citet{fre13} predicted that around 15 - 20\% of main-sequence FGK field stars host small planets (0.8 - 1.25 R$_\oplus$) with orbital $<$ 85 d. This ratio is consistent with those reported for solar twins \citep{mel09,ram09,ram10}. If we assume a terrestrial planet fraction of 15\%, and all terrestrial planets imprint the chemical signatures on to the hosts, then we would estimate that $\approx$ 2.4 programme stars should be unusual in their chemical composition in our sample. Given the small number statistics, the null result is consistent with the prediction according to the terrestrial planet frequency in the field stars. Tentatively, we conclude that our analysis thus provides an independent constraint upon the fraction of open cluster stars that might host terrestrial planets.

\begin{figure}
\centering
\includegraphics[width=\columnwidth]{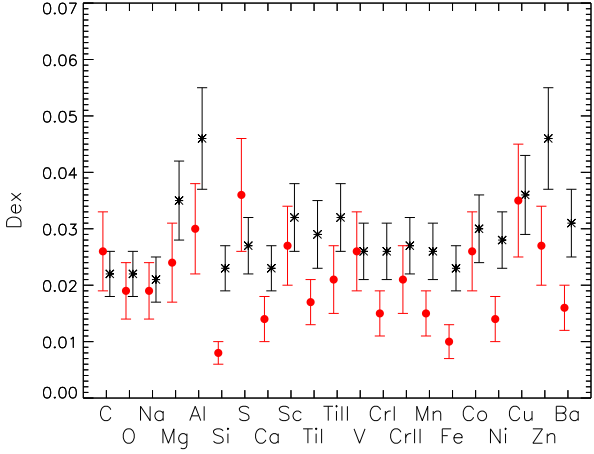}
\caption{Observed abundance dispersions (black asterisk) and average abundance errors ($<$$\sigma\Delta^{\rm X}$$>$, red circles) for all species in our sample. These results were obtained when using the reference star HD 25825.}
\label{fig7}
\end{figure}

\begin{figure}
\centering
\includegraphics[width=\columnwidth]{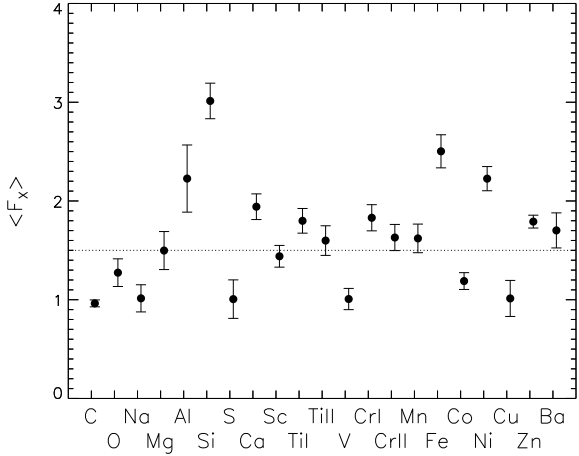}
\caption{The mean $F_X$ for each species using all reference stars in our sample. The dashed line locates at $F_X$ = 1.5, which means that the abundance dispersion is 1.5 times larger than the average measurement error.}
\label{fig7.5}
\end{figure}

\begin{figure}
\centering
\includegraphics[width=\columnwidth]{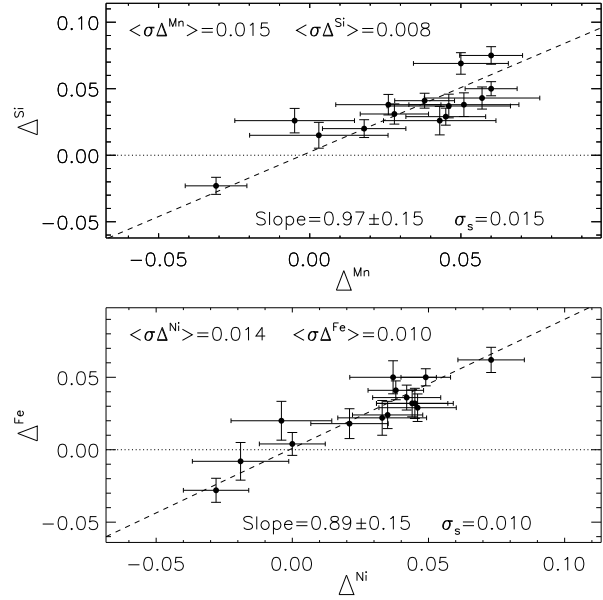}
\caption{Upper panel: $\Delta^{\rm Si}$ versus $\Delta^{\rm Mn}$; lower panel: $\Delta^{\rm Fe}$ versus $\Delta^{\rm Ni}$, for the programme stars when using the reference star HD 25825. The dashed lines represent linear fits. $\sigma_s$ is the dispersion about the linear fit. We write the average abundance errors in x-axis and y-axis ($<$$\sigma\Delta^{\rm X}$$>$ and $<$$\sigma\Delta^{\rm Y}$$>$, respectively).}
\label{fig8}
\end{figure}

\subsection{Star-to-star abundance variations among the Hyades stars}

In order to detect any chemical signature of planet formation, we have achieved the highest chemical abundance precision ever obtained in an open cluster. With this unique data set, we can study chemical homogeneity among the Hyades open cluster. We plot the average abundance error $<$$\sigma\Delta^{\rm X}$$>$, and the measured abundance dispersion (standard deviation), for all elements in Figure \ref{fig7}. The main result from this figure is that we have achieved very high precision in the differential chemical abundances of our programme stars by applying the strictly line-by-line analysis technique. The lowest average abundance error is for Si ($<\sigma\Delta^{\rm Si}>$ = 0.008 dex) and the highest values is for S ($<\sigma\Delta^{\rm S}>$ = 0.036 dex). Previous studies of the Hyades achieved typical abundance errors of $\sim$ 0.05 - 0.06 dex but reaching as low as $\sim$ 0.03 - 0.04 dex for some elements (P03; \citealp{de06}). Another important aspect to note in Figure \ref{fig7} is that the measured dispersions for many elements (12 out of 19) are considerably larger than the average abundance errors by a factor of $\sim$ 1.5 - 2. We note that the real abundance errors for C, S and Cu could be overestimated due to the lower S/N around the spectral region of these elements. In Table 6, we write the total abundance variation as well as the standard deviation, and the average abundance error for each element, using HD 25825 as the reference star. We find that the average abundance errors are smaller than the observed abundance dispersions for most elements. This is the first evidence that the Hyades is chemically inhomogeneous. An alternative explanation, however, is that we have underestimated the errors. 

In order to quantify the level of chemical inhomogeneity, we define the fraction $F_X$ which represents the ratio of abundance dispersion to the errors. A value of $F_X$ = 1 means that the abundance dispersion is equal to the measurement error while $F_X$ = 2 means that the abundance dispersion is twice the measurement error. For a given element using a particular reference star, we performed 10,000 realizations in which we draw random numbers from the observed abundance dispersion distribution and from the distribution of average uncertainties. For a given element, we repeated this exercise using each reference star in turn. From this, we derived the mean $F_X$ for each element using all reference stars and show the results in Figure \ref{fig7.5}. This plot further confirms the results presented in Figure \ref{fig7} using HD 25825 as the reference star.

We searched for correlations between different elements in the differential chemical abundances ($\Delta^{\rm X}$ versus $\Delta^{\rm Y}$) to further investigate the abundance variations in our Hyades stars. In Figure \ref{fig8}, we plot two examples of $\Delta^{\rm X}$ versus $\Delta^{\rm Y}$ ($\Delta^{\rm Si}$ versus $\Delta^{\rm Mn}$ in the upper panel, and $\Delta^{\rm Fe}$ versus $\Delta^{\rm Ni}$ in the lower panel, respectively). We applied a linear least-squares fit to the data, taking into account errors in both variables and in each panel we show the slope and corresponding uncertainty. Consideration of the slopes and uncertainties of the linear fits reveals that while the amplitude may be small, there are statistically significant, positive correlations between these elements for our programme stars. The significance level of the linear fits are 6$\sigma$ for both combinations. While underestimating the errors could explain the results presented in Figures \ref{fig7} and \ref{fig7.5}, it is highly unlikely that correlations of such high statistical significance between pairs of elements would arise from underestimating the errors.

\begin{figure*}
\centering
\includegraphics[width=0.98\hsize]{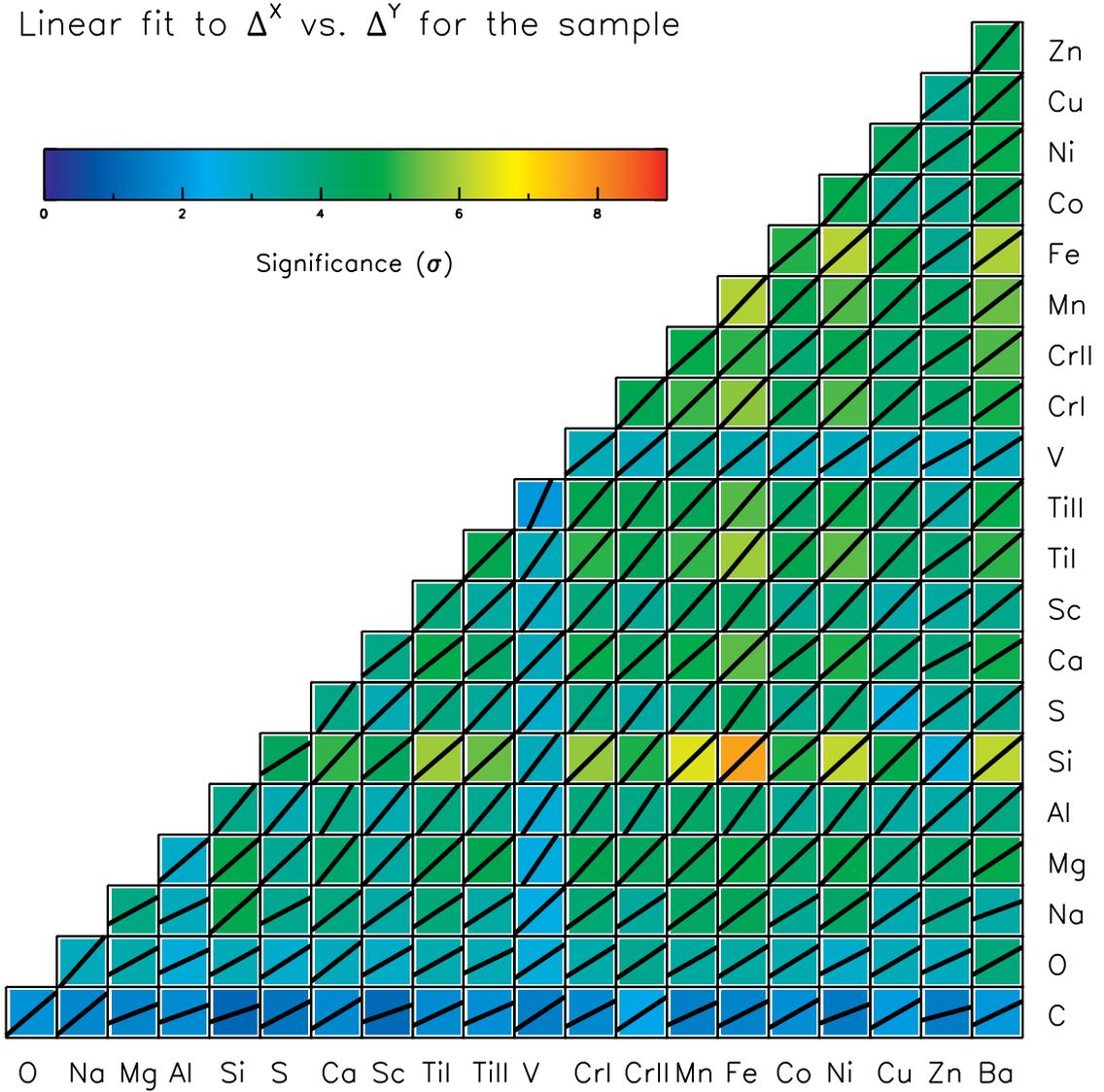}
\caption{Linear least-squares fit to $\Delta^{\rm X}$ versus $\Delta^{\rm Y}$, for all the combinations of species. The dimensions of the x-axis and y-axis are unity. The colour bar indicates the signifiance of the gradients. These results were obtained when using the reference star HD 25825.}
\label{fig9}
\end{figure*}

We then show $\Delta^{\rm X}$ versus $\Delta^{\rm Y}$, for every possible combination of species in Figure \ref{fig9}. The dimensions of the x-axis and y-axis are unity, such that a slope of gradient 1.0 would be represented by a straight line from the lower-left corner to the upper-right corner and a slope of gradient 0.0 would be a horizontal line. The different colours in Figure \ref{fig9} indicate corresponding significance levels, which are based on the slopes and the uncertainties. The gradients are always positive and most of them ($\approx$ 90\% of the pairs) have significance $>$ 2.5$\sigma$. We note that the correlations with Si are of the highest statistical significance, probably because Si has the lowest error. We conclude that there are positive correlations, of high statistical significance, between at least 90\% of pairs of elemental abundances. Similar results have been reported for the globular cluster NGC 6752 by \citet{dy13}. We interpret the ubiquitous positive correlations, often of high statistical significance, between $\Delta^{\rm X}$ and $\Delta^{\rm Y}$ as further indication of a genuine abundance dispersion in the Hyades.

We then calculated the intrinsic abundance scatter for each element in our sample using the selected reference star (HD 25825) in the following manner. For each element, we adopt a Gaussian distribution of width = 0.001 dex (which is the initial guess of the intrinsic abundance scatter) and randomly draw numbers from this distribution (one for each star). Then we add in quadrature another random number drawn from a Gaussian distribution of width corresponding to the error for that element in that programme star. We repeat this process 1000 times for all stars and measure the average produced value. We iterate the whole procedure by increasing the guess of the intrinsic abundance scatter by 0.001 dex until we find the "real intrinsic abundance scatter" which reproduces the observed abundance dispersion. Table 6 lists the values of intrinsic abundance scatter for each element in our sample, using HD 25825 as the reference star. We note that the average value of the intrinsic abundance scatter is 0.021 $\pm$ 0.003 dex ($\sigma$ = 0.010).

\begin{table}
\caption{The total abundance variation as well as the standard deviation (abundance dispersion), the average abundance error, and the intrinsic abundance scatter for each element in our sample, using HD 25825 as the reference star.}
\centering
\begin{tabular}{@{}lcccc@{}}
\hline
Species & Total & Standard & Average & Intrinsic \\
 & variation & deviation & error & scatter \\
\hline
C    & 0.080 & 0.022 & 0.026 & 0.003 \\
O    & 0.080 & 0.022 & 0.019 & 0.014 \\
Na   & 0.079 & 0.021 & 0.019 & 0.011 \\
Mg   & 0.117 & 0.035 & 0.024 & 0.028 \\
Al   & 0.175 & 0.046 & 0.030 & 0.039 \\
Si   & 0.098 & 0.023 & 0.008 & 0.024 \\
S    & 0.088 & 0.027 & 0.036 & 0.001 \\
Ca   & 0.078 & 0.023 & 0.014 & 0.019 \\
Sc   & 0.103 & 0.032 & 0.027 & 0.021 \\
TiI  & 0.099 & 0.029 & 0.017 & 0.026 \\
TiII & 0.110 & 0.032 & 0.021 & 0.027 \\
V    & 0.080 & 0.026 & 0.026 & 0.013 \\
CrI  & 0.091 & 0.026 & 0.015 & 0.023 \\
CrII & 0.105 & 0.027 & 0.021 & 0.019 \\
Mn   & 0.091 & 0.026 & 0.015 & 0.022 \\
Fe   & 0.090 & 0.023 & 0.010 & 0.023 \\
Co   & 0.090 & 0.030 & 0.026 & 0.019 \\
Ni   & 0.101 & 0.028 & 0.014 & 0.027 \\
Cu   & 0.127 & 0.036 & 0.035 & 0.017 \\
Zn   & 0.169 & 0.046 & 0.027 & 0.042 \\
Ba   & 0.114 & 0.031 & 0.016 & 0.030 \\
\hline
\end{tabular}
\end{table}

\subsection{Detailed examination of systematic errors}

We have made several tests to check for possible systematic errors which might affect our results and describe them below.

(a) Systematic errors in EW measurements

\begin{figure}
\centering
\includegraphics[width=\columnwidth]{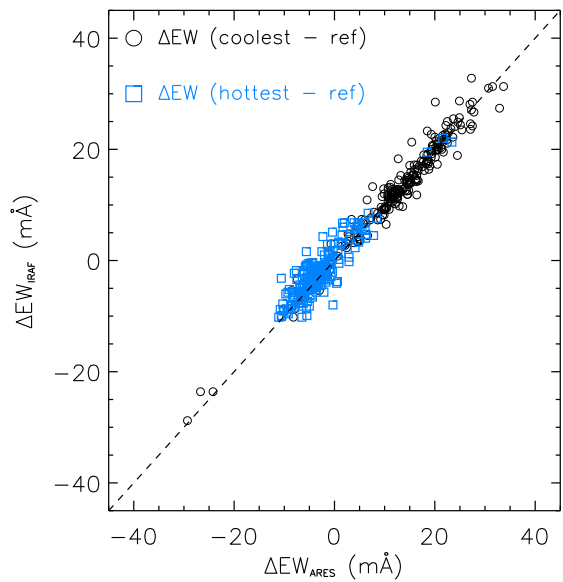}
\caption{Differential EWs of spectral lines of the coolest and the warmest sample stars (HD 27282: black circles and HD 28205: blue rectangles, respectively) with respect to the reference star HD 25825, measured by ARES (x-axis) and IRAF (y-axis). The black dotted line represents the one-to-one relation.}
\label{fig:dew}
\end{figure}

Rather than manually measuring the spectral lines with the careful placement of the continuum at the same level in similar stars, an automatic code, ARES \citep{sou07} was used to measure EWs of the adopted lines in this work. ARES performs a local normalization around each spectral line, which might introduce small systematic differences in the adopted continuum between different lines. Therefore we present a test to compare the differential EWs measured by ARES with that measured manually using IRAF. We measured the differential EWs of spectral lines of the coolest and the warmest sample stars (HD 27282 and HD 28205, respectively) with respect to the reference star HD25825. Figure \ref{fig:dew} shows the comparison results. The measurements of differential EWs with ARES and IRAF clearly show one-to-one relations, which indicate that no systematic errors are induced due to the use of ARES. We made a further test by restricting only strong lines ($>$ 80 m\AA), while the comparison results are similar as shown in Figure \ref{fig:dew}, which demonstrate the ARES does not necessarily introduce systematic errors in the EWs as a function of effective temperature, as well as microturbulent velocity.

b) Errors in effective temperature

\begin{figure}
\centering
\includegraphics[width=\columnwidth]{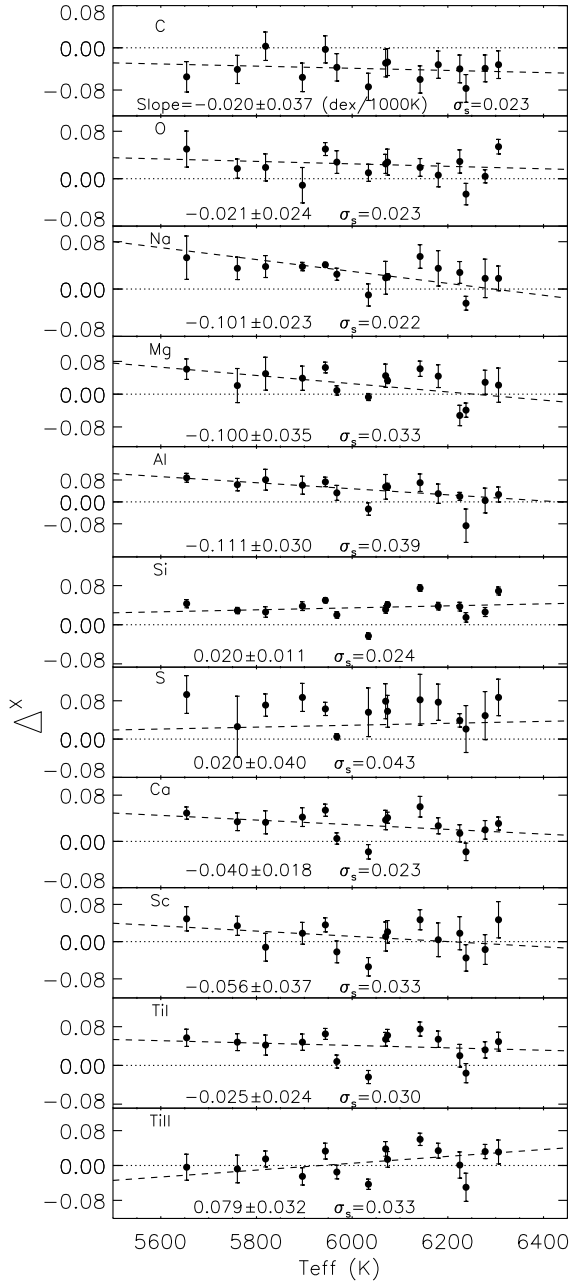}
\caption{$\Delta^{\rm X}$ versus T$_{\rm eff}$ for C, O, Na, Mg, Al, Si, S, Ca, Sc, TiI, and TiII for the programme stars when using the reference star HD 25825. The black dashed lines represent the linear fit to the data. $\sigma_s$ is the dispersion about the linear fit.}
\label{fig10}
\end{figure}

\begin{figure}
\centering
\includegraphics[width=\columnwidth]{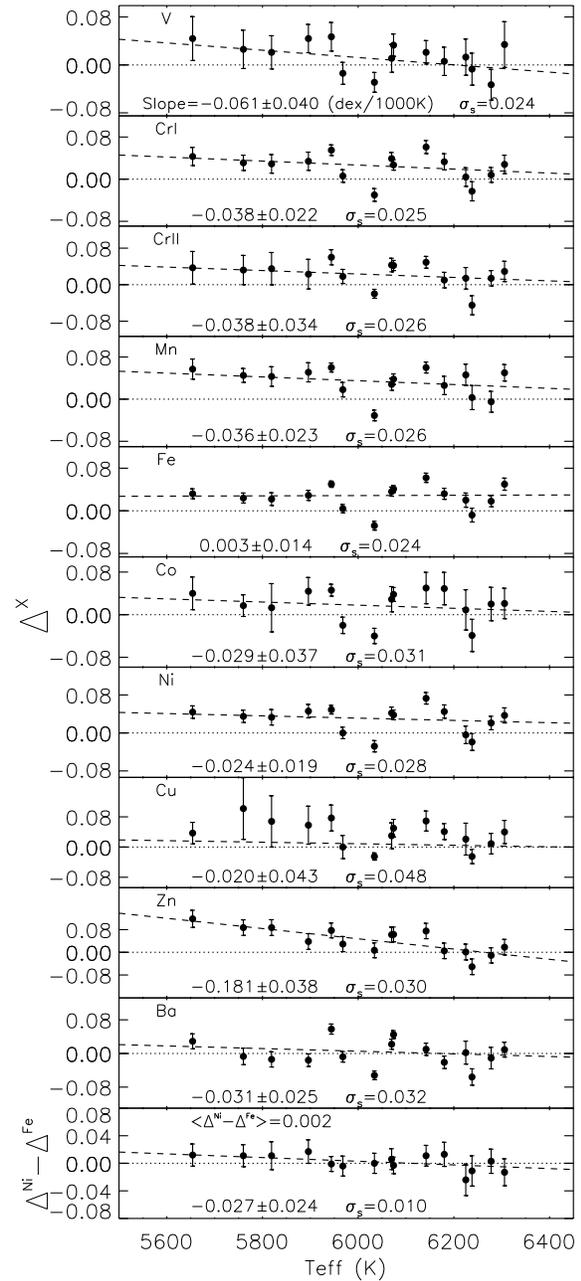}
\caption{Same as Figure \ref{fig10} but for V, CrI, CrII, Mn, Fe, Co, Ni, Cu, Zn, and Ba, as well as ($\Delta^{\rm Ni}$ $-$$\Delta^{\rm Fe}$) versus T$_{\rm eff}$ for the programme stars when using the reference star HD 25825. The black dashed lines represent the linear fit to the data. $\sigma_s$ is the dispersion about the linear fit.}
\label{fig10.5}
\end{figure}

\begin{figure*}
\centering
\includegraphics[width=0.98\hsize]{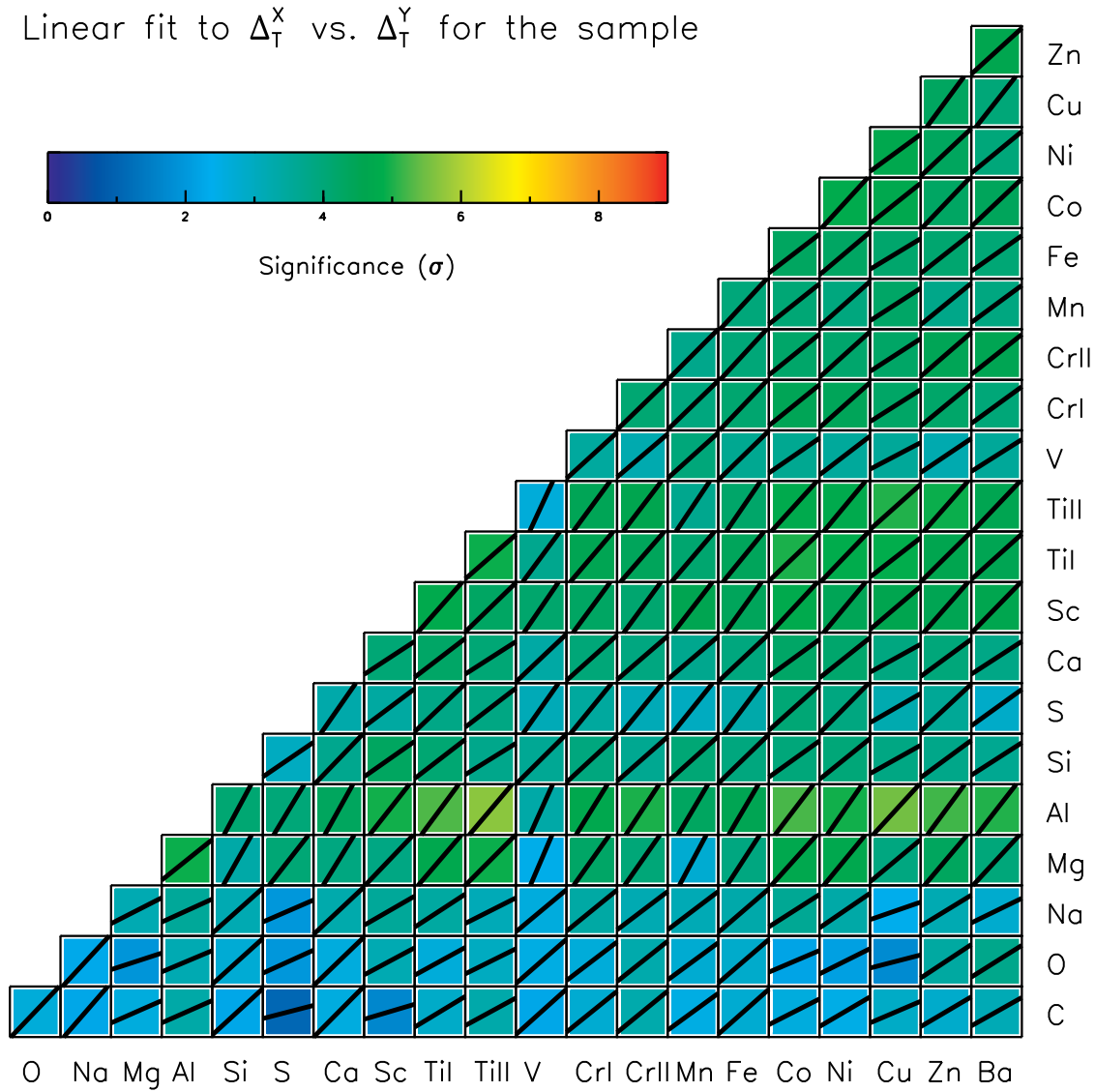}
\caption{Same as Figure \ref{fig9} but the abundance trends with T$_{\rm eff}$ have been removed. These results were obtained when using the reference star HD 25825.}
\label{fig11}
\end{figure*}

\begin{figure}
\centering
\includegraphics[width=0.98\hsize]{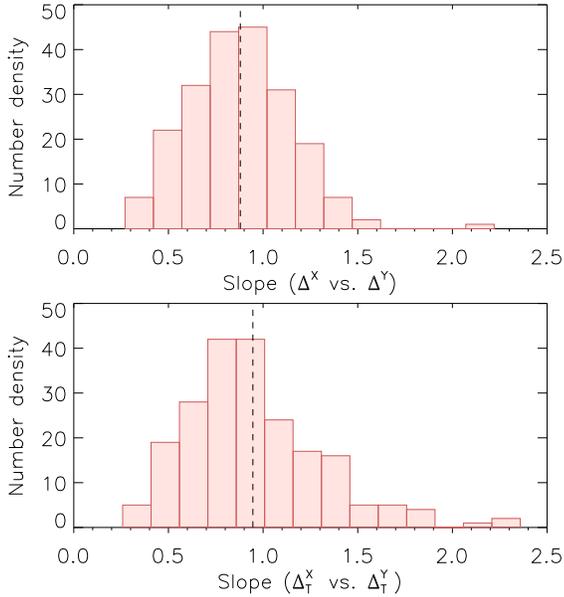}
\caption{The distribution of the slopes for the linear least-squares fits to $\Delta$X versus $\Delta$Y, for all the combinations of species without removing the T$_{\rm eff}$ trends (upper panel) and with the T$_{\rm eff}$ trends have been removed (lower panel). The dashed vertical lines represent the location of the mean value of $\Delta^{\rm X}$ versus $\Delta^{\rm Y}$ slopes.}
\label{fig11.5}
\end{figure}

We plot $\Delta^{\rm X}$ versus T$_{\rm eff}$ for all the elements in Figures \ref{fig10} and \ref{fig10.5}. These two plots suggest that there may be trends between differential chemical abundances and T$_{\rm eff}$. Since the total range in T$_{\rm eff}$ is large ($\sim$ 660 K), we tentatively attribute these trends to differential NLTE or 3D effects (e.g., \citealp{asp05}). For example, Zn, that is the worst case, seem to have the right effect for the 472nm line according to \citet{tak05} using the Delta-1 model or Delta-2 model, which could introduce $\sim$ 0.07 dex difference in $\Delta$[Zn/H] for the coolest and the warmest sample stars. Therefore we need to explore whether or not our results (abundance trends between $\Delta^{\rm X}$ versus $\Delta^{\rm Y}$) change if we remove the abundance trends with T$_{\rm eff}$. We removed the abundance trends with T$_{\rm eff}$ in the following way. We defined a new quantity, $\Delta^{\rm X}_{\rm T}$, which is the difference between $\Delta^{\rm X}$ and the value of the linear fit to the data at the T$_{\rm eff}$ of the programme star. Then we examine the trends between $\Delta^{\rm X}_{\rm T}$ and $\Delta^{\rm Y}_{\rm T}$ in Figure \ref{fig11}. This figure is similar as Figure \ref{fig9} but we have removed the abundance trends with T$_{\rm eff}$. The results are essentially unchanged for all pairs of elements: at least 90\% of pairs of elements show positive correlations of the similar significance as before. We note that none of the elements have slopes that differ by 2$\sigma$. In addition, we show the distribution of all the slopes from Figure \ref{fig9} and Figure \ref{fig11} in Figure \ref{fig11.5}. The mean value of the slopes without removing the T$_{\rm eff}$ trends is 0.88 $\pm$ 0.02 ($\sigma$ = 0.27), while the mean value of the slopes with the T$_{\rm eff}$ trends have been removed is 0.95 $\pm$ 0.02 ($\sigma$ = 0.36). This test increases our confidence that our results are not an artefact of systematic errors in terms of T$_{\rm eff}$.

However, we note that for most elements, the programme stars with T$_{\rm eff}$ $>$ 5900 K show larger abundance variations when compared to the programme stars with T$_{\rm eff}$ $\leq$ 5900 K. This could be related to the thin convection zones of those stars with T$_{\rm eff}$ $>$ 5900 K since it is easier to imprint abundance anomalies. Another possibility is diffusion. \citet{on14} detected tentative variations in the open cluster M67 that could be due to atomic diffusion, albeit M67 is much older than the Hyades. \citet{geb10} reported large abundance variations (e.g., $\sim$ 0.2 dex) in A and F stars due to diffusion, although their F stars are hotter than our sample stars by $\sim$ 1000 K. We plot ($\Delta^{\rm Ni}$ $-$ $\Delta^{\rm Fe}$) versus T$_{\rm eff}$ in Figure \ref{fig10.5} (bottom panel). We find that the abundance difference between these two elements is almost zero while the predicted abundance difference from the diffusion model should be $\sim$ $+$0.2 dex (1.45 M$_\odot$ case, \citealp{ric00}). In this scenario, the hotter and more massive stars should have higher Ni to Fe ratios than the cooler and less massive stars. We do not detect such a trend and therefore we do not find evidence in our sample for diffusion effects.

Earlier we noted that systematic errors cancel in a differential analysis. Previous analyses usually spanned a small range in T$_{\rm eff}$ $\pm$ 100 K (e.g., \citealp{mel09,ram14a}). Here, our programme stars span $\sim$ 300 K in T$_{\rm eff}$. Examination of Figures \ref{fig10} and \ref{fig10.5} indicate that there are no significant ($>$ 2.5$\sigma$) systematic trends between abundance and T$_{\rm eff}$ for most elements except for Na, Al, and Zn, which would suggest that the systematic errors cancel over this range of T$_{\rm eff}$.

c) Effects of T$_{\rm eff}$, $\log g$ and $\xi_{\rm t}$ error vectors

Next we seek to understand whether individual errors in T$_{\rm eff}$, $\log g$ and $\xi_{\rm t}$ could induce abundance trends between $\Delta^{\rm X}$ versus $\Delta^{\rm Y}$ that mimic our results. The tests are presented in the following manner. We kept the reference star (HD 25825) fixed. Starting with T$_{\rm eff}$, we computed new abundances by randomly changing T$_{\rm eff}$ according to the uncertainty ($\sigma$T$_{\rm eff}$) for each programme star. Assuming the data all lie at [0.0,0.0] in $\Delta^{\rm X}$ versus $\Delta^{\rm Y}$, we can then generate a new plot in which the fit to these data effectively represent the "T$_{\rm eff}$ error vector". We can then quantify whether errors in T$_{\rm eff}$ can mimic the measurements. Error vectors can be obtained for $\log g$ and $\xi_{\rm t}$, by applying a similar approach using the uncertainties $\sigma$$\log g$ and $\sigma$$\xi_{\rm t}$, respectively. The underlying hypothesis we were testing was whether the distribution in $\Delta^{\rm X}$ versus $\Delta^{\rm Y}$ is a $\delta$ function centred at the zero-point and that the observed distribution could be explained entirely by errors in T$_{\rm eff}$ or $\log g$ or $\xi_{\rm t}$.

\begin{figure}
\centering
\includegraphics[width=\columnwidth]{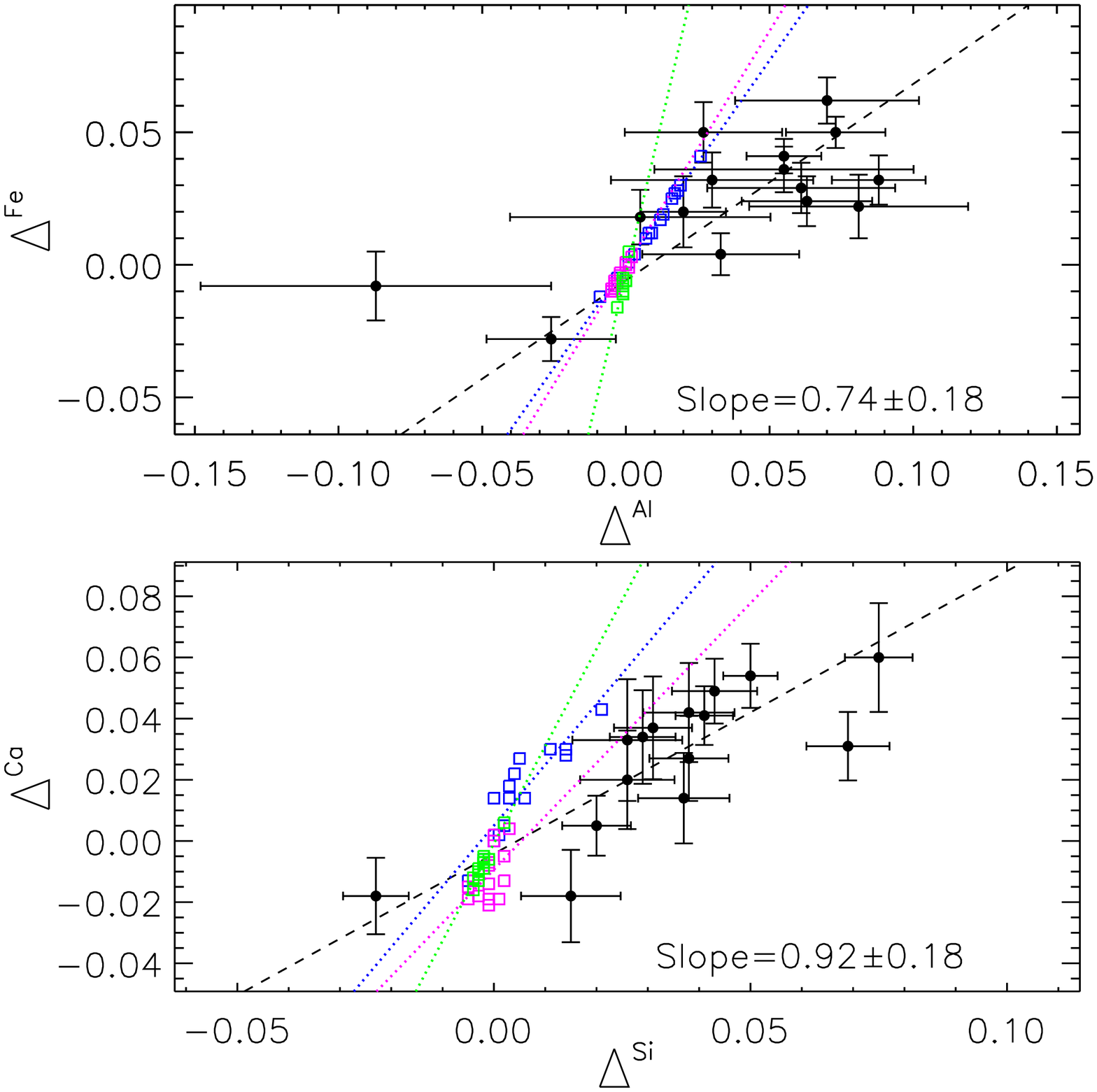}
\caption{Upper panel: $\Delta^{\rm Fe}$ versus $\Delta^{\rm Al}$; lower panel: $\Delta^{\rm Ca}$ versus $\Delta^{\rm Si}$, for the programme stars when using the reference star HD 25825. The black dashed lines represent the linear least-squares fit to the data. The blue, magenta and green dashed lines represent the error vectors of T$_{\rm eff}$, $\log g$ and $\xi_{\rm t}$, respectively.}
\label{fig12}
\end{figure}

We plot two examples of $\Delta^{\rm X}$ versus $\Delta^{\rm Y}$ ($\Delta^{\rm Fe}$ versus $\Delta^{\rm Al}$ in the upper panel, and $\Delta^{\rm Ca}$ versus $\Delta^{\rm Si}$ in the lower panel, respectively) with error vectors of T$_{\rm eff}$, $\log g$ and $\xi_{\rm t}$ (blue, magenta and green dashed lines, respectively) in Figure \ref{fig12}. It is clear that the errors in T$_{\rm eff}$ or $\log g$ or $\xi_{\rm t}$ alone can not fully explain the observed trends in $\Delta^{\rm Fe}$ versus $\Delta^{\rm Al}$ and $\Delta^{\rm Ca}$ versus $\Delta^{\rm Si}$ since the error vectors are not aligned with the data, and as discussed, the magnitude of the errors is far smaller than the observed dispersions. We applied this test to all the pairs of elements. The fraction of instances in which the error vectors of T$_{\rm eff}$, $\log g$ and $\xi_{\rm t}$ are in agreement with the observed trends including uncertainties are 25\%, 12\% and 20\%, respectively. This indicates that the variations of these three stellar parameters can not fully explain the positive correlations for the vast majority ($>$ 75\%) of differential elemental abundances shown in Figure \ref{fig9}. We also checked our results by multiplying the errors in T$_{\rm eff}$, $\log g$ and $\xi_{\rm t}$ by a factor of 2, 3 and 5 and applied the similar manner described above. Naturally this can only increase the amplitude of the error while the direction of the error vector remains unchanged. This test reinforces that our main results are not likely due to systematic errors in stellar parameters.

d) Effects of stellar activity

To investigate the potential effects of stellar activity on our results, we computed the chromospheric activity index log R'$_{\rm HK}$ as follows. We measured the fluxes in the cores of the Ca II H and K lines using 1 \AA \ triangular passbands. Pseudo-continuum fluxes were measured using 20 \AA \ bandpasses in the continuum at 3901 and 4001 \AA. We can thus measure the instrumental $S_{\rm inst}$ index (see, e.g., \citealp{wri04}) from our spectra. We found a linear relationship between our $S_{\rm inst}$ index and $S_{\rm Duncan}$ (values published in \citealp{dun91}):
\begin{equation}
S_{\rm Duncan} = 0.023(\pm0.057) + 0.082(\pm0.180)S_{\rm inst}
\end{equation}
Thus we are able to transform our $S_{\rm inst}$ values into a standard Mount Wilson S index scale ($S_{\rm MW}$). B $-$ V colours listed in the Hipparcos Catalog \citep{esa97} were then employed to transform  $S_{\rm MW}$ into log R'$_{\rm HK}$ using equations from \citet{mid82} and \citet{noy84}. Our measurements of log R'$_{\rm HK}$ show good agreement with previously published values of common Hyades stars \citep{dun91,pau02}. When compared to the results from \citet{pau02} (hereafter P02), the mean difference (our values $-$ P02 values) is $-$0.06 $\pm$ 0.06. Thus, our log R'$_{\rm HK}$ values have errors of the order $\sim$ 0.06 and there is little time variation of this activity index in the programme stars between our observations and those of P02.

\begin{figure}
\centering
\includegraphics[width=\columnwidth]{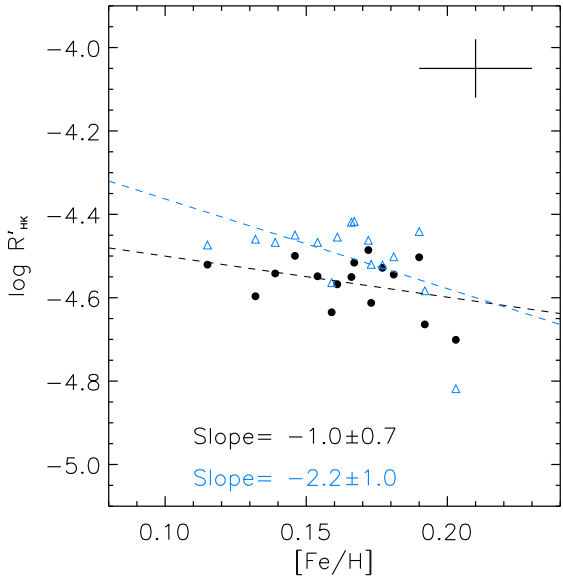}
\caption{The stellar chromospheric activity index log R'$_{\rm HK}$ versus derived [Fe/H] for our Hyades stars. The black circles represent the index measured based on our spectra, while the blue triangles represent the index taken from P02. The black dashed line and the blue dashed line represent the linear least-squares fit to our data and P02 data, respectively.}
\label{fig13}
\end{figure}

We would like to check if the abundance variations and the observed positive correlations of elemental abundances are due to the effects of stellar chromospheric activity. Figure \ref{fig13} shows the stellar activity index log R'$_{\rm HK}$ versus [Fe/H] for our sample. We did not find any clear relation between the stellar activity index and our derived [Fe/H], no matter our results or P02 results were adopted. Instead, they are distributed more or less randomly. We made this test for all the other elements and found that none of them show correlations with $>$ 2.5$\sigma$ significance. Therefore, the observed abundance variations and correlations of elemental abundances can not be physically attributed to the effects of stellar activity.

\subsection{Possible explanations for an intrinsic abundance spread}

Our results offer the first clear evidence that the Hyades open cluster is chemically inhomogeneous at the $\approx$ 0.02 dex level. Chemical inhomogeneity at this level can only be detected when the measurement uncertainties are extremely small, as in our study. Here we discuss several potential scenarios, which could explain the observed abundance variations and positive correlations between $\Delta^{\rm X}$ versus $\Delta^{\rm Y}$ in the Hyades stars. We note that in principle, the possible explanations do not have to be able to create inhomogeneities in all chemical abundances, but only on those that have abundance dispersions above the measurement errors (as in, e.g., Figure \ref{fig7.5}).

a) Inhomogeneous chemical evolution in the proto-cluster environment

In this scenario, we assume that the abundance variations and correlations are due to chemical inhomogeneities in the proto-cluster environment. Our Hyades data indicate that all elements are positively correlated, regardless of their nucleosynthetic origin. For example, the $\alpha$-element Ca is positively correlated with the Fe-peak element Ni as well as with the $s$-process element Ba. The correlations between light, $\alpha$-, Fe-peak and neutron-capture elements demand contributions from a variety of nucleosynthetic sources, and it would seem unlikely that this is the explanation. Similarly, GCE would not affect all elements equally such that they evolve in lock-step (e.g., \citealp{kob11}).

b) Supernova ejection in the proto-cluster cloud

Of particular interest is the fact that some of the elements which exhibit star-to-star variations and correlations are synthesized in massive stars that die as core collapse supernovae (SNe II). A typical SNe II from a 15 M$_\odot$ produces $\sim$ 10$^{-1}$ M$_\odot$ of Fe \citep{ww95}. We assume that the mass for the giant molecular cloud from which the Hyades was formed was $\sim$ 800 - 1600 M$_\odot$ \citep{wei92,kb02}. The mass fraction of Fe from that SNe II in such a cloud will be $\sim$ (1.25 - 0.63) $\times$ 10$^{-4}$. The Fe content of the Sun is $\approx$ 1.5 $\times$ 10$^{-3}$ M$_\odot$ \citep{asp09}. If we assume [Fe/H] $\approx$ 0.1 dex for our case, the corresponding Fe content will be $\sim$ 1.9 $\times$ 10$^{-3}$ M$_\odot$. Thus we can estimate the change in Fe abundance, produced by such a typical SNe II, will be $\approx$ 0.02 - 0.04 dex, which is comparable with the intrinsic abundance scatter in Fe abundance in our sample ($\approx$ 0.023 dex). Therefore, one SNe II can account for the change in Fe abundance in the Hyades.

The supernova timescale (t$_{\rm SN}$) is $\approx$ 3 Myr and we would expect the open clusters not to be fully homogeneous if they were assembled on timescales longer than the supernova timescale and all gas is expelled once the SNe explodes. Since no clear separation in timescales between chemical homogenization and the star formation, the time required for turbulent mixing to smooth out the proto-cluster gas cloud might be longer than $\sim$ 3 Myr for the Hyades open cluster, if the hypothesis is true. We note that the main constraint derived here is limited not by the constraint on the abundance spread, but instead by whether a core collapse supernovae of a massive star is likely to have occurred and to have polluted the star-forming gas where the Hyades open cluster formed. A further problem of this scenario is that the supernova ejecta can not produce all elements to reveal the abundance variations seen in our results.

c) Dilution with metal-poor gas

One possibility is that metal-poor gas might pollute the molecular star-forming cloud. Theoretical simulations suggested that the gas and dust in star-forming clouds can be very well mixed \citep{fk14}, which would lead to an abundance scatter $\sim$ 0.01 - 0.05 dex. However, when we are able to achieve a precision level of $\approx$ 0.02 dex in our strictly line-by-line differential abundance analysis, we note that the open cluster Hyades shows the inhomogeneities for many elements since the abundance dispersions are $\sim$ 0.025 - 0.045 dex, a factor of 1.5 - 2 larger than the predicted errors, as shown in Figure \ref{fig7}, leading to an intrinsic abundance scatter of $\sim$ 0.02 dex. According to \citet{fk14}, the turbulent mixing during cloud assembly would happen when the star formation efficiency reaches $\sim$ 30 \% for the clusters with mass $\sim$ 10$^3$ M$_\odot$. Therefore, the pollution of metal-poor gas should happen before within $\sim$ 3 Myr. In addition, our results also provide constrains on the intrinsic abundance dispersion in the molecular cloud where the Hyades formed. Using the prediction from \citet{fk14}, the proto-cluster cloud would have abundance scatter $\sim$ 5 times higher than the abundance scatter in the Hyades, which would lead to $\sim$ 0.1 dex scatter in the gas abundances.

If we assume that the most metal-rich stars represent the "true" abundance of the Hyades, then we can estimate how much dilution is needed to produce the most metal-poor Hyades objects. In the limit that the diluting material is metal-free, then a mixture of eight parts "true" Hyades material to one part diluting material would result in a decrease in [X/H] of 0.04 dex, for all elements. In the more likely event that the diluting material is not metal free, then the mixture shifts in favour of the diluting material. For example, if the diluting material half that of the "true" Hyades composition, then a mixture of 3.5 parts Hyades material to one part diluting material would result in a decrease of 0.05 dex in [X/H]. Theoretical simulations are needed to examine whether such dilution is dynamically plausible. We note that pollution of metal-rich gas is another possibility since the same arguments can apply.

\section{Conclusion}

We have studied the Hyades, a benchmark open cluster, to investigate whether we can detect chemical signatures of planet formation. We analysed 16 solar-type stars in the Hyades based on high resolution, high signal-to-noise ratio (S/N $\approx$ 350 - 400 per pixel) spectra obtained from the McDonald 2.7m telescope, allowing us to achieve very high precision in stellar parameters and differential chemical abundances with uncertainties as small as 0.008 dex for our programme stars.

We did not find any significant correlations in abundance with condensation temperature for our Hyades stars in the \citet{mel09} scenario. We demonstrated that the observed abundance dispersions in our Hyades stars are a factor for $\approx$ 1.5 - 2 larger than the average measurement errors for most elements, and that there is an intrinsic abundance dispersion of 0.021 $\pm$ 0.003 dex ($\sigma$ = 0.010) in the Hyades open cluster. The differential chemical abundances of at least 90\% of pairs of elements have positive correlations with high statistical significance, which strengthens our statement that the Hyades is chemically inhomogeneous. Removing the abundance trends with T$_{\rm eff}$ do not alter our results. We recall that the abundance trends with T$_{\rm eff}$ might be due to modelling errors. We do not find evidence in our data for atomic diffusion effects in the Hyades. Tests on the error vectors of the stellar atmospheric parameters indicate that $>$ 75\% of the positive correlations between $\Delta^{\rm X}$ and $\Delta^{\rm Y}$ can not be explained by changing the stellar parameters systematically. Additionally and importantly, these results persist regardless of the choice of reference star, i.e., the results are independent of the reference star. We note that the chemical inhomogeneities are not due to the planet effects, considering the lack of T$_{\rm cond}$ trends in our sample. The possible scenarios of these abundance variations include: (a) inhomogeneous chemical evolution in the proto-cluster environment, (b) supernova ejection in the proto-cluster cloud, (c) pollution of metal-poor, or metal-rich, gas before complete mixing of the proto-cluster cloud.

Our detailed differential abundance analysis for the Hyades stars provides significant constraints upon the chemical homogeneity of open clusters and a challenge to the current view of Galactic archeology, in terms of "chemical tagging". The Hyades is the first, and thus far only, open cluster to which we have applied high precision chemical abundance techniques. By extension, it may be that other (perhaps all)  open clusters are similarly chemically inhomogeneous. Clearly it is important to extend this type of analysis to additional open clusters to identify chemical signatures of planet formation and/or chemical inhomogeneity.

\section*{Acknowledgments}
This work has been supported by the Australian Research Council (grants FL110100012, FT140100554 and DP120100991). JM thanks support by FAPESP (2012/24392-2). We thank the detailed and helpful suggestions provided by the referee of this paper.

\section*{SUPPLEMENTARY MATERIAL}

Additional Supporting Information may be found in the online version of this article:

\noindent\\
Table A1. Atomic-line data, as well as the EW measurements, adopted for the abundance analysis.



\bsp

\label{lastpage}


\begin{thebibliography}{99}

\bibitem[\protect\citeauthoryear{Adibekyan et al.}{2014}]{adi14} Adibekyan V.Z., Gonz\'alez Hern\'andez J.I., Delgado Mena E., et al. 2014, A\&A, 564, L15
\bibitem[\protect\citeauthoryear{Amarsi et al.}{2016}]{ama16} Amarsi A.M., Asplund M., Collet R., Leenaarts J., 2016, MNRAS, 455, 3735
\bibitem[\protect\citeauthoryear{Asplund}{2005}]{asp05} Asplund M., 2005, ARA\&A, 43, 481
\bibitem[\protect\citeauthoryear{Asplund et al.}{2009}]{asp09} Asplund M., Grevesse N., Sauval A.J., Scott P., 2009, ARA\&A, 47, 481
\bibitem[\protect\citeauthoryear{Bensby et al.}{2005}]{ben05} Bensby T., Feltzing S., Lundstr\"om I., Ilyin I., 2005, A\&A, 433, 185
\bibitem[\protect\citeauthoryear{Bensby, Feltzing \& Oey}{2014}]{ben14} Bensby T., Feltzing S., Oey M.S., 2014, A\&A, 562, A71
\bibitem[\protect\citeauthoryear{Biazzo et al.}{2015}]{bia15} Biazzo K., Gratton R., Desidera S., et al. 2015, A\&A, 583, 135
\bibitem[\protect\citeauthoryear{Blanco Cuaresma et al.}{2015}]{bc15} Blanco-Cuaresma S., Soubiran C., Heiter U., et al. 2015, A\&A, 577, A47
\bibitem[\protect\citeauthoryear{Bland-Hawthorn \& Freeman}{2004}]{bf04} Bland-Hawthorn J., Freeman K., 2004, PASA, 21, 110
\bibitem[\protect\citeauthoryear{Bland-Hawthorn et al.}{2010a}]{bla10a} Bland-Hawthorn J., Karlsson T., Sharma S., et al. 2010a, ApJ, 721, 582
\bibitem[\protect\citeauthoryear{Bland-Hawthorn, Krumholz \& Freeman}{2010b}]{bla10b} Bland-Hawthorn J., Krumholz M.R., Freeman K., 2010b, ApJ, 713, 166
\bibitem[\protect\citeauthoryear{Boesgaard \& Tripicco}{1986}]{bt86} Boesgaard A.M., Tripicco M.J., 1986, ApJ, 302, L49
\bibitem[\protect\citeauthoryear{Bovy}{2016}]{bo16} Bovy J., 2016, ApJ, 817, 49
\bibitem[\protect\citeauthoryear{Brandt \& Huang}{2015}]{bh15} Brandt T.D., Huang C.X., 2015, ApJ, 807, 58
\bibitem[\protect\citeauthoryear{Brucalassi et al.}{2014}]{bru14} Brucalassi A., Pasquini L., Saglia R., et al. 2014, A\&A, 561, 9
\bibitem[\protect\citeauthoryear{Carrera \& Pancino}{2011}]{cp11} Carrera R., Pancino E., 2011, A\&A, 535, A30
\bibitem[\protect\citeauthoryear{Castelli \& Kurucz}{2003}]{cas03} Castelli F., Kurucz R.L., 2003, IAU Symposium, 210, 20
\bibitem[\protect\citeauthoryear{Chambers}{2010}]{cha10} Chambers J.E., 2010, ApJ, 724, 92
\bibitem[\protect\citeauthoryear{Chiappini, Matteucci \& Romano}{2001}]{chi01} Chiappini C., Matteucci F., Romano D., 2001, ApJ, 554, 1044
\bibitem[\protect\citeauthoryear{Cochran, Hatzes \& Paulson}{2002}]{coc02} Cochran W.D., Hatzes A.P., Paulson D.B., 2002, AJ, 124, 565
\bibitem[\protect\citeauthoryear{De Silva et al.}{2006}]{de06} De Silva G.M., Sneden C., Paulson D.B., et al. 2006, AJ, 131, 455
\bibitem[\protect\citeauthoryear{De Silva et al.}{2007}]{de07} De Silva G.M., Freeman K., Asplund M., et al. 2007, AJ, 133, 1161
\bibitem[\protect\citeauthoryear{De Silva et al.}{2015}]{de15} De Silva G.M., Freeman K., Bland-Hawthorn J., et al. 2015, MNRAS, 449, 2604
\bibitem[\protect\citeauthoryear{Duncan et al.}{1991}]{dun91} Duncan D.K., Vaughan A.H., Wilson O.C., et al. 1991, ApJS, 76, 383
\bibitem[\protect\citeauthoryear{Dutra-Ferreira et al.}{2016}]{dut16} Dultra-Ferreira L., Pasquini L., Smiljanic R., et al. 2016, A\&A, 585, A75
\bibitem[\protect\citeauthoryear{Edvardsson et al.}{1993}]{edv93} Edvardsson B., Andersen J., Gustafsson B., et al. 1993, A\&A, 275, 101
\bibitem[\protect\citeauthoryear{Epstein et al.}{2010}]{eps10} Epstein C.R., Johnson J.A., Dong S., et al. 2010, ApJ, 709, 447
\bibitem[\protect\citeauthoryear{ESA}{1997}]{esa97} ESA 1997, ESA SP-1200: The Hipparcos and Tycho Catalogues. ESA, Noordwijk
\bibitem[\protect\citeauthoryear{Feng \& Krumholz}{2014}]{fk14} Feng Y., Krumholz M.R., 2014, Nature, 513, 523
\bibitem[\protect\citeauthoryear{Freeman \& Bland-Hawthorn}{2002}]{fb02} Freeman K., Bland-Hawthorn J., 2002, ARA\&A, 40, 487
\bibitem[\protect\citeauthoryear{Fressin et al.}{2013}]{fre13} Fressin F., Torres G., Charbonneau D., et al. 2013, ApJ, 766, 81
\bibitem[\protect\citeauthoryear{Friel \& Boesgaard}{1992}]{fb92} Friel E.D., Boesgaard A.M., 1992, ApJ, 387, 170
\bibitem[\protect\citeauthoryear{Friel et al.}{2014}]{fri14} Friel E.D., Donati P., Bragaglia A., et al. 2014, A\&A, 563, A117
\bibitem[\protect\citeauthoryear{Gebran et al.}{2010}]{geb10} Gebran M., Vick M., Monier R., Fossati L., 2010, A\&A, 523, A71
\bibitem[\protect\citeauthoryear{Gonz\'alez Hern\'andez et al.}{2010}]{gh10} Gonz\'alez Hern\'andez J.I., Israelian G., Santos N.C., et al. 2010, ApJ, 720, 1592
\bibitem[\protect\citeauthoryear{Grevesse et al.}{2015}]{gre15} Grevesse N., Scott P., Asplund M., Sauval A.J., 2015, A\&A, 573, 27
\bibitem[\protect\citeauthoryear{Griffin et al.}{1988}]{gri88} Griffin R.F., Griffin R.E.M., Gunn J.E., Zimmerman B.A., 1988, AJ, 96, 172
\bibitem[\protect\citeauthoryear{Kobayashi, Karakas \& Umeda}{2011}]{kob11} Kobayashi C., Karakas A.I., Umeda H., 2011, MNRAS, 414, 3231
\bibitem[\protect\citeauthoryear{Kraft}{1994}]{kra94} Kraft R.P., 1994, PASP, 106, 553
\bibitem[\protect\citeauthoryear{Kroupa \& Boily}{2002}]{kb02} Kroupa P., Boily C.M., 2002, MNRAS, 336, 1188
\bibitem[\protect\citeauthoryear{Kubryk, Prantzos \& Athanassoula}{2015}]{kub15} Kubryk M., Prantzos N., Athanassoula E., 2015, A\&A, 580, A126
\bibitem[\protect\citeauthoryear{Kurucz \& Bell}{1995}]{kur95} Kurucz R., Bell B., 1995, Kurucz CD-ROM, NO. 23, Harvard-Smithsonian Centre for Astrophysics
\bibitem[\protect\citeauthoryear{Lada \& Lada}{2003}]{lad03} Lada C.J., Lada E.A., 2003, ARA\&A, 41, 57
\bibitem[\protect\citeauthoryear{Liu et al.}{2014}]{liu14} Liu F., Asplund M., Ram\'irez I, Yong D, Mel\'endez J., 2014, MNRAS, 442, L51
\bibitem[\protect\citeauthoryear{Liu et al.}{2016}]{liu16} Liu F., Yong D., Asplund M., et al. 2016, MNRAS, 456, 2636
\bibitem[\protect\citeauthoryear{Lodders}{2003}]{lod03} Lodders K., 2003, ApJ, 591, 1220
\bibitem[\protect\citeauthoryear{Lovis \& Mayor}{2007}]{lm07} Lovis C., Mayor M., 2007, A\&A, 472, 657
\bibitem[\protect\citeauthoryear{Maderak et al.}{2013}]{mad13} Maderak R.M., Deliyannis C.P., King J.R., Cummings J.D., 2013, AJ, 146, 143
\bibitem[\protect\citeauthoryear{Mann et al.}{2015}]{man15} Mann A.W., Gaidos E., Mace G.N., et al. 2015, arXiv:1512.00483
\bibitem[\protect\citeauthoryear{Masseron \& Gilmore}{2015}]{mg15} Masseron T., Gilmore G., 2015, MNRAS, 453, 1855
\bibitem[\protect\citeauthoryear{Meibom et al.}{2013}]{mei13} Meibom S., Torres G., Fressin F., et al. 2013, Nature, 499, 55
\bibitem[\protect\citeauthoryear{Mel\'endez et al.}{2009}]{mel09} Mel\'endez J., Asplund M., Gustafsson B., Yong D., 2009, ApJ, 704, L66
\bibitem[\protect\citeauthoryear{Mel\'endez et al.}{2012}]{mel12} Mel\'endez J., Bergemann M., Cohen J.G., et al. 2012, A\&A, 543, A29
\bibitem[\protect\citeauthoryear{Middelkoop}{1982}]{mid82} Middelkoop F., 1982, A\&A, 107, 31
\bibitem[\protect\citeauthoryear{Monroe et al.}{2013}]{mon13} Monroe T.R., Mel\'endez J., Ram\'irez I., et al. 2013, ApJ, 774, L32
\bibitem[\protect\citeauthoryear{Morel \& Micela}{2004}]{mm04} Morel T., Micela G., 2004, A\&A, 423, 677
\bibitem[\protect\citeauthoryear{Neves et al.}{2009}]{nev09} Neves V., Santos N.C., Sousa S.G., Correia A.C.M., Israelian G., 2009, A\&A, 497, 563
\bibitem[\protect\citeauthoryear{Nissen}{2015}]{nis15} Nissen P.E., 2015, A\&A, 579, A52
\bibitem[\protect\citeauthoryear{Noyes et al.}{1984}]{noy84} Noyes R.W., Hartmann L.W., Baliunas S.L., et al. 1984, ApJ, 279, 763
\bibitem[\protect\citeauthoryear{\"Onehag et al.}{2011}]{on11} \"Onehag A., Korn A., Gustafsson B., Stempels E., VandenBerg D.A., 2011, A\&A, 528, A85
\bibitem[\protect\citeauthoryear{\"Onehag, Gustafsson \& Korn}{2014}]{on14} \"Onehag A., Gustafsson B., Korn A., 2014, A\&A, 562, A102
\bibitem[\protect\citeauthoryear{Paulson et al.}{2002}]{pau02} Paulson D.B., Saar S.H., Cochran W.D., Hatzes A.P., 2002, AJ, 124, 572
\bibitem[\protect\citeauthoryear{Paulson, Sneden \& Cochran}{2003}]{pau03} Paulson D.B., Sneden C., Cochran W.D., 2003, AJ, 125, 3185
\bibitem[\protect\citeauthoryear{Paulson, Cochran \& Hatzes}{2004}]{pau04} Paulson D.B., Cochran W.D., Hatzes A.P., 2004, AJ, 127, 3579
\bibitem[\protect\citeauthoryear{Perryman et al.}{1998}]{per98} Perryman M.A.C., Brown A.G.A., Lebreton Y., et al. 1998, A\&A, 331, 81
\bibitem[\protect\citeauthoryear{Pichardo et al.}{2012}]{pic12} Pichardo B., Moreno E., Allen C., et al. 2012, AJ, 143, 73
\bibitem[\protect\citeauthoryear{Quinn et al.}{2012}]{qui12} Quinn S.N., White R.J., Latham D.W., et al. 2012, ApJ, 756, L33
\bibitem[\protect\citeauthoryear{Quinn et al.}{2014}]{qui14} Quinn S.N., White R.J., Latham D.W., et al. 2014, ApJ, 787, 27
\bibitem[\protect\citeauthoryear{Ram\'irez, Mel\'endez \& Asplund}{2009}]{ram09} Ram\'irez I., Mel\'endez J., Asplund M., 2009, A\&A, 580, L17
\bibitem[\protect\citeauthoryear{Ram\'irez et al.}{2010}]{ram10} Ram\'riez I., Asplund M., Baumann P., Mel\'endez J., Bensby T., 2010, A\&A, 521, A33
\bibitem[\protect\citeauthoryear{Ram\'irez et al.}{2014a}]{ram14a} Ram\'irez I., Mel\'endez J., Bean J., et al. 2014, A\&A, 572, A48
\bibitem[\protect\citeauthoryear{Ram\'irez et al.}{2014b}]{ram14b} Ram\'irez., Mel\'endez J., Asplund M., 2014, A\&A, 561, A7
\bibitem[\protect\citeauthoryear{Ram\'irez et al.}{2015}]{ram15} Ram\'irez I., Khanal S., Aleo P., et al. 2015, ApJ, 808, 13
\bibitem[\protect\citeauthoryear{Randich et al.}{2005}]{ran05} Randich S., Bragaglia A., Pastori L., et al. 2005, The Messenger, 121, 18
\bibitem[\protect\citeauthoryear{Richer, Michaud \& Turcotte}{2000}]{ric00} Richer J., Michaud G., Turcotte S., 2000, ApJ, 529, 338
\bibitem[\protect\citeauthoryear{Saffe, Flores \& Buccino}{2015}]{saf15} Saffe C., Flores M., Buccino A., 2015, A\&A, 582, A17
\bibitem[\protect\citeauthoryear{Sato et al.}{2007}]{sa07} Sato B., Izumiura H., Toyota E., et al. 2007, ApJ, 661, 527
\bibitem[\protect\citeauthoryear{Scott et al.}{2015a}]{sco15a} Scott P., Grevesse N., Asplund M., et al. 2015, A\&A, 573, 25
\bibitem[\protect\citeauthoryear{Scott et al.}{2015b}]{sco15b} Scott P., Asplund M., Grevesse N., et al. 2015, A\&A, 573, 26
\bibitem[\protect\citeauthoryear{Smiljanic et al.}{2010}]{smi10} Smiljanic R., Pasquini L., Charbonnel C., Lagarde N., 2010, A\&A, 510, A50
\bibitem[\protect\citeauthoryear{Sneden}{1973}]{sne73} Sneden C., 1973, ApJ, 184, 839
\bibitem[\protect\citeauthoryear{Sousa et al.}{2007}]{sou07} Sousa S.G., Santos N.C., Israelian G., Mayor M., Monteiro M.J.P.F.G., 2007, A\&A, 469, 783
\bibitem[\protect\citeauthoryear{Spina, Mel\'endez \& Ram\'irez}{2016}]{spi16} Spina L., Mel\'endez J., Ram\'irez I., 2016, A\&A, 585, A152
\bibitem[\protect\citeauthoryear{Takeda et al.}{2005}]{tak05} Takeda Y., Hashimoto O., Taguchi H., et al. 2005, PASJ, 57, 751
\bibitem[\protect\citeauthoryear{Ting et al.}{2012}]{ti12} Ting Y.S., De Silva G.M., Freeman K., et al. 2012, MNRAS, 427, 882
\bibitem[\protect\citeauthoryear{Tucci Maia, Mel\'endez \& Ram\'irez}{2014}]{tm14} Tucci Maia M., Mel\'endez J., Ram\'irez I., 2014, ApJ, 790, L25
\bibitem[\protect\citeauthoryear{Tull et al.}{1995}]{tu95} Tull R.G., MacQueen P.J., Sneden C., Lambert D.L., 1995, PASP, 107, 251
\bibitem[\protect\citeauthoryear{Weidemann et al.}{1992}]{wei92} Weidemann V., Jordan S., Iben I.J., Casertano S., 1992, AJ, 104, 1876
\bibitem[\protect\citeauthoryear{Woosley \& Weaver}{1995}]{ww95} Woosley S.E., Weaver T.A., 1995, ApJS, 101, 181
\bibitem[\protect\citeauthoryear{Wright et al.}{2004}]{wri04} Wright J.T., Marcy G.W., Butler R.P., Vogt S.S., 2004, ApJS, 152, 261
\bibitem[\protect\citeauthoryear{Wright et al.}{2012}]{wri12} Wright J.T., Marcy G.W., Howard A.W., et al. 2012, ApJ, 753, 160
\bibitem[\protect\citeauthoryear{Yong et al.}{2013}]{dy13} Yong D., Mel\'endez J., Grundahl F., et al. 2013, MNRAS, 434, 3542

\end{thebibliography}
\end{document}